\documentclass{article}[11pt]
\usepackage{graphicx} 
\usepackage[caption=false]{subfig} 
\usepackage{authblk} 

\usepackage{xcolor}
\usepackage{natbib}
\usepackage{amsmath}
\usepackage{amssymb}
\usepackage{bbm}
\usepackage[figuresright]{rotating}
\usepackage{multicol,ragged2e}
\usepackage{setspace}
\usepackage{algorithm}
\usepackage{algpseudocode}
\usepackage{booktabs}
\usepackage{threeparttable}
\usepackage{multirow}
\usepackage{array}
\usepackage{url} 
\usepackage{xr-hyper}
\usepackage{xr} 

\usepackage{tikz}
\usetikzlibrary{shapes}
\usetikzlibrary{arrows}
\definecolor{revised}{rgb}{0,0,0.9}

\makeatletter
\newcommand*{\addFileDependency}[1]{
  \typeout{(#1)}
  \@addtofilelist{#1}
  \IfFileExists{#1}{}{\typeout{No file #1.}}
}
\makeatother

\newif\ifarXiv
\arXivtrue 

\newcommand*{\myexternaldocument}[1]{%
    \externaldocument{#1}%
    \addFileDependency{#1.tex}%
    \addFileDependency{#1.aux}%
}
\myexternaldocument{suppArxiv}

 \newcounter{acounter}
\newtheorem{theorem}{Theorem}

\newcommand{\nl}{\newline}
\newcommand{\pl}{\parallel}
\newcommand{\openr}{\hbox{${\rm I\kern-.2em R}$}}
\newcommand{\openn}{\hbox{${\rm I\kern-.2em N}$}}

\title{A cautionary note for plasmode simulation studies in the setting of causal inference}
\date{}
\date{}
\author[1,2,*]{Pamela~A. Shaw}
\author[3]{Susan Gruber}
\author[1,2]{Brian~D. Williamson}
\author[4]{Rishi Desai}
\author[1,2]{Susan~M. Shortreed}
\author[1]{Chloe Krakauer}
\author[1,2]{Jennifer~C. Nelson}
\author[5]{Mark~J. van der Laan}
\affil[1]{Biostatistics Division, Kaiser Permanente Washington Health Research Institute, Seattle, WA, USA}
\affil[2]{Department of Biostatistics, University of Washington, Seattle, WA, USA}
\affil[3]{TL Revolution, LLC, Cambridge, MA, USA}
\affil[4]{Division of Pharmacoepidemiology and Pharmacoeconomics, Department of Medicine, Brigham and Women’s Hospital, Harvard Medical School, Boston, MA, USA}
\affil[5]{Department of Biostatistics, School of Public Health, University of California at Berkeley, Berkeley, CA, USA}
\affil[*]{Corresponding Author: Biostatistics Division, Kaiser Permanente Washington Health Research Institute, 1730 Minor Ave Ste 1360, Seattle, WA 98101. Email: pamela.a.shaw@kp.org}

\begin{document}
  
\maketitle
\newpage
\begin{abstract} 
Plasmode simulation has become an important tool for evaluating the operating characteristics of different statistical methods in complex settings, such as pharmacoepidemiological studies of treatment effectiveness using electronic health records (EHR) data.  These studies provide insight into how estimator performance is impacted by challenges including rare events, small sample size, etc., that can indicate which among a set of methods performs best in a real-world dataset. Plasmode simulation combines data resampled from a real-world dataset with synthetic data to generate a known truth for an estimand in realistic data. There are different potential plasmode strategies currently in use.  We compare two popular plasmode simulation frameworks. We provide numerical evidence and a theoretical result, which shows that one of these frameworks can cause certain estimators to incorrectly appear overly biased with lower than nominal confidence interval coverage. Detailed simulation studies using both synthetic and real-world EHR data demonstrate that these pitfalls remain at large sample sizes and when analyzing data from a randomized controlled trial. We conclude with guidance for the choice of a plasmode simulation approach that maintains good theoretical properties to allow a fair evaluation of statistical methods while also maintaining the desired similarity to real data.

\end{abstract}

\section{Introduction}
Simulation studies are typically used to empirically evaluate the relative performance (e.g., bias, efficiency) of estimation methods \citep{morris2019using,koehler2009assessment}. One concern is that simplistic computer-generated data structures may lead to performance that is different than what would be expected in real-world data. This limitation motivated a simulation technique known as ``plasmode simulation'', in which one or more key elements of the data, such as the treatment effect, is computer-generated, but others are generated by bootstrap sampling real-world data. 

 Several authors have used plasmode simulation to select the best performing method to handle missingness in confounders in electronic health records (EHR) data \citep{weberpals2024principled, weberpals2024high, williamson2024assessing, getz2023performance}. It has been used to evaluate machine learning methods to build prediction models \citep{hafermann2022using,karim2024can}, to study the performance of different propensity score approaches and other causal methods to estimate treatment effects \citep{conover2021propensity,ress2024comparing,rudolph2023simulation}, and to compare statistical methods in a variety settings  \citep{alfaras2021empirical,iyassu2024identification,duchesneau2022timing}.

Despite the frequency of use, formal mathematical justification to guide the proper application of plasmode simulation has not been fully considered.   One of the first uses of the term ``plasmode'' to describe data where key elements are known to fit some true underlying model was by \citet{cattell1967general}.  Decades later \citet{gadbury2008evaluating} define plasmode  ``to describe datasets that are derived from real data but for which some truth is known,'' and demonstrate the utility of plasmode simulation in high-dimensional settings. 

\citet{franklin_plasmode_2014} described a now popular plasmode simulation procedure to evaluate estimation methodologies for a treatment exposure estimand in the context of longitudinal health care databases. Their method specifies bootstrap sampling covariates and treatment together, then generating the outcome conditional on resampled data. They suggest this \textit{Sample Treatment} framework preserves covariate-treatment relationships, while still allowing the investigator to know the true (simulated) causal effect.  
An alternate plasmode simulation approach is to bootstrap sample only the covariates ($W$), then simulate treatment ($A$) conditional on the covariates. In keeping with the plasmode philosophy, one can generate $A$ from a propensity score model trained on the actual data, then generate the outcome, $Y$, conditional on $W$ and then generated $A$.  Many papers have used this \textit{Generate Treatment} plasmode framework in analyses of pharmacoepidemiologic data \citep{wang2018relative,conover2021propensity,franklin2017comparing}. In a recent tutorial paper on plasmode simulation, \citet{schreck_statistical_2024} discuss that in addition to sampling covariates, one can choose to also simulate artificial covariates from a parametric model, but do not go into details as to when one would want to do so. 

We compare the common \textit{Sample Treatment} and \textit{Generate Treatment} plasmode frameworks and show only the latter approach can be reliably used to evaluate the statistical performance of many typical casual inference methods for evaluating a point treatment effect. We begin by defining a set of casual estimands and common statistical methods whose performance will be evaluated with plasmode simulation. We then describe the two plasmode frameworks and discuss why only the Generate Treatment framework will satisfy the mathematical conditions necessary to guarantee the desirable analytical properties when evaluating performance for several estimators relying on the propensity score. Numerical studies are performed to evaluate the two plasmode simulation procedures for a variety of fully synthetic data settings. We conduct a series of plasmode simulation studies using real-world data to further illustrate the superior performance of the proposed Generate Treatment framework. We conclude with a summary of lessons learned regarding plasmode simulation to evaluate statistical methods.


\section{Motivating setting}

In observational studies, a principal concern when estimating a treatment effect is confounding bias. 
We show that the problematic issues that arise when doing Sample Treatment plasmode simulations are relevant quite generally for common estimators of causal estimands.  A key issue is that not all estimators are affected in the same manner by this plasmode approach, which means the perceived relative performance of the estimators can be misleading. In this section, we formally define a set of treatment effect estimands and estimators to be compared using the two plasmode simulation frameworks. 


\subsection{Notation and estimands}

Consider data with $n$ i.i.d. observations $O = (W, A, Y)$, $W$ a vector of baseline covariates, $A$ a binary treatment indicator, and outcome $Y$. We define potential outcome $Y(a)$ for treatment exposure $A=a$.  Let $\mu_a = E[Y(a)]$. We define the marginal ATE = $\mu_1 - \mu_0$ and RR $= \mu_1 / \mu_0$. We also consider the coefficient for treatment $A$ in a logistic regression model for $Y$ adjusted for $W$, namely the log conditional odds ratio (logcOR) for $A$, a common estimand for binary outcomes.

\subsection{Estimators} \label{sec:estimators}
We seek to evaluate the performance of algorithms that require correct specification of the propensity score (PS) for consistency, as well as those that do not. 

We consider the following common estimators for the ATE: 
\begin{itemize}
    \item[-] \textbf{Propensity score matching (Match)}. This algorithm uses the generalized full optimal matching algorithm with replacement \citep{hansen2004full,savje2021generalized} to generate weights for each observation suitable for evaluating the ATE. The outcome model for $ E(Y|A)$ is then estimated using a weighted, unadjusted linear regression of the outcome on treatment.
    \item[-] \textbf{Inverse probability of treatment weighting (IPTW)}.  Observation weights are stabilized using the marginal treatment probability \citep{robins2000marginal} and bounded to be no more than $\sqrt{n} \ln(n)/5$ \citep{gruber2022data}. The outcome model $E(Y|A)$ is then estimated using a weighted, unadjusted linear regression of the outcome on treatment.
    \item[-] \textbf{Doubly robust targeted maximum likelihood estimation (TMLE)}. The TMLE \citep{vanderlaan2006targeted} is fit using the correctly specified working models for the treatment propensity and outcome, bounding the treatment assignment probabilities by $5 / (\sqrt{n} \ln(n)).$
    \item[-] \textbf{Generalized linear model, correctly specified (glmCM)}. The outcome model $ E(Y|A,W)$ is fit using the correctly specified regression model.
    \item[-] \textbf{Generalized linear model, adjusted for propensity score (glmPS)}. The outcome model is fit regressing $Y$ on $A$ and the PS fit from a correctly specified model $ E(A|W)$.
\end{itemize}
For each estimator, predicted counterfactual outcomes  are obtained when $A$ is set to 0 and to 1 to estimate the ATE (or RR).

A typical estimation approach for the logcOR relies on a marginal structural model (MSM) for the outcome. We define a parametric logistic MSM that includes $A$ and a subset of potential confounders, $\mathbf{\widetilde{W}}$. The MSM logcOR is estimated by weighted maximum likelihood using IPTW. Further details are provided in the numerical studies section below.



\section{Plasmode simulation}
We define plasmode simulation as a data generation mechanism (DGM) for which at least part of the constructed data is generated by bootstrap sampling a source dataset $\mathcal{D}$. Commonly the source dataset is from  a real-world setting and the goal of plasmode simulation is to create datasets with similar  structure to the source data but for which there is a known truth. We assume that an outcome $Y^{\#}$ will be generated from $\mathcal{D}$ so that the treatment estimand of interest (ATE, RR or logcOR) has known value. 

For the Sample Treatment plasmode approach, both the treatment and covariates $(A,W)$ are bootstrap sampled from the source data $\mathcal{D}$ and then $Y^{\#}$ is stochastically generated conditional on $(A,W)$ according to an assumed model, defined as $f_Y(A,W, U_Y)$, where $U_Y$ is an exogenous error term.  For the Generate Treatment plasmode approach, we define a propensity score model $P(A|W)=f_A(W,U_A)$, which may be a fitted model on the source data.   We then bootstrap sample only $W$ from $\mathcal{D}$, generate $A^{\#}$ from $P(A|W)$, and then generate the outcome, $Y^{\#}$, conditional on $(W,A^{\#})$, from  $f_Y(A, W,U_Y)$ as before. The two approaches are summarized in Table~\ref{tab:resampling}.


\section{Theoretical problems for the Sample Treatment framework} \label{sec:Maintheory}

A frequent goal of a plasmode simulation is to evaluate whether a particular method can accurately and efficiently estimate an estimand of interest. To this end, the statistical performance of an estimator based on data generated by plasmode simulation needs to resemble the statistical performance of the estimator based on the actual data distribution.  This does not necessarily mean that plasmode simulations are drawn from  an accurate estimator of the data distribution (e.g., think outcome blind simulations), but qualitatively it should be highly informative for comparing estimators and evaluating performance of a given estimator. We provide theoretical justification for why samples generated from the Sample Treatment framework will not produce accurate estimates of estimator performance, such as bias or confidence interval coverage.

The root of the problem is that the positivity assumption required for identifying a causal estimand,  $P(A = a| W)>0$ for all $a$ and observed $W$ in the data, is violated under the Sample Treatment framework. Under this plasmode approach,  every time $W_i$ is sampled, the associated value for $A_i$ is fixed at some $a$, its value in the original data for subject $i$; thus, the probability of receiving treatment $a$ given $W_i$ is 1. Importantly, this positivity violation occurs even when there is no violation in the source population. 

This induced violation of the positivity assumption under the Sample Treatment framework implies that estimators relying on outcome regression for consistency (e.g., parametric G-computation, such as glmCM) are fully reliant on extrapolation for the treatment/covariate combinations missed by the Sample Treatment algorithm, whereas estimators relying on propensity score estimation (IPTW, glmPS or Match) will end up having non-negligible bias in the plasmode samples, even when the propensity score model is correctly specified. We show bias is of order $n^{-1/2}$, which means it is non-negligible with respect to the SE of the estimator, even as sample size converges to infinity, which affects confidence interval coverage. 

For example, given data $O_i=(Y_i,A_i,W_i) \sim P_0$, suppose the target estimand is the average outcome $Y$ for treatment $A=1$,  namely $\Psi(P_0)= E_0[E_0(Y|A=1,W)]$. For an i.i.d. sample $O_1,\ldots,O_n$ of size $n$ drawn from $P_0$, whose empirical distribution we denote with $P_n$, consider the simple IPTW-estimator $\hat{\Psi}(P_n)=n^{-1}\sum_{i=1}^n A_iY_i/\hat{g}(P_n)$, where $\hat{g}(P_n)$ is a maximum likelihood estimator of the true propensity score $\bar{g}_0(W)=P_0(A=1\mid W)$ based on a correctly specified parametric model. This IPTW estimator is a consistent, asymptotically linear estimator of $\Psi(P_0)$, with negligible bias of order $1/n$. 

Consider now the Sample Treatment data distribution ${\bf P}_n$ that draws $(W,A)$ from $P_n$, and then draws $Y$ given $(W,A)$ from a specified conditional distribution, such as the distribution of $Y$ given $(W,A)$ under $P_0$. Under such a data distribution, one would view the true estimand as $\psi_{n}=\Psi({\bf P}_n)=n^{-1} \Sigma_{i=1}^n E(Y|W_i=w_i,A=1)$ (strictly speaking $\Psi({\bf P}_n)$ is not well defined due to the positivity issue, which makes the conditional mean improper for observations $W_i$ with $A_i=0$, and that is the core problem). We show that under the Sample Treatment data distribution $\mathbf{P}_n$, the IPTW estimand $\hat{\psi}(\mathbf{P}_n)$ (i.e., the target of the IPTW estimator it ends up fitting when given an infinite sample from ${\bf P}_n$)  is given by $n^{-1}\Sigma_{i=1}^n A_i/\hat{g}(P_n) E(Y\mid W=W_i,A=1)$, which is not equal to $\psi_n$, and the difference equals $B_n= \hat{\psi}(\mathbf{P}_n)- \psi_n =  n^{-1} \Sigma_{i=1}^n \left [ \frac{\{A_i-\bar{g}_n(W)\} E(Y\mid W=W_i,A=1)}{\bar{g}_n(W)} \right ]$, where $\bar{g}_n=\hat{g}(P_n)$. We show that this bias $B_n$  approximates zero only at rate $n^{-1/2}$, but not faster. That means the IPTW estimator acts as an inconsistent estimator of $\psi_n$ in the plasmode samples from ${\bf P}_n$. Even though the bias $B_n$ goes to 0 at rate $n^{-1/2}$, that bias is of the same order as the SE and thus non-negligible for coverage, wrongly suggesting that the IPTW estimator is an overly biased estimator.

This bias stems from the difference between $P_0$ and $\textbf{P}_n$ with respect to the treatment assignment mechanism, which does not go away asymptotically due to the induced positivity violation. Conversely, the data distribution for the Generate Treatment plasmode approach $\mathbf{\tilde{P}}_n$ avoids these theoretical problems by assigning positive probability on $A=1$ and $A=0$ given any possible drawn $W$.  A detailed mathematical argument is presented in Supplemental Appendix \ref{sec:theory}.

\section{Numerical studies}\label{sec:SimCity}

\subsection{Overview}

We demonstrate how the choice of the plasmode simulation framework impacts the assessment of estimator performance. Four simulation scenarios utilize completely synthetic data, varying the outcome model, treatment assignment mechanism, rarity of a binary outcome, and treatment estimand. The fifth scenario mimics a real-world plasmode simulation,  using a source dataset from the Kaiser Permanente Washington (KPWA) EHR system \citep{williamson2024assessing}. We examine estimator performance for the estimators outlined in Section \ref{sec:estimators}, as well as an unadjusted estimator, which provides a sense of the degree of confounding. The unadjusted estimate (Unadj) is similar to the glmCM and glmPS methods, except treatment is the only covariate in the outcome model.

For each DGM we conduct 100,000 Monte Carlo replications and evaluate the mean percent (\%) bias, empirical standard error (SE), root mean-squared error (RMSE) and absolute bias:SE ratio. Estimators under study are asymptotically regular (with $\sqrt{n}$-convergence rate); thus, for consistent estimators, we expect the bias:SE to be proportional to $1/\sqrt{n}$. 
For asymptotically biased estimators (e.g., Unadj in most scenarios), bias:SE increases with sample size  because SE shrinks to 0 while bias does not. 
We considered $n= 100$, $1000$, and $10,000$ in Scenarios 1-4.  For the real-data Scenario 5, $n=10,000$.

\subsection{Synthetic data simulation set-up}
\subsubsection{Scenarios 1 and 2: Continuous and binary outcomes}
 Scenarios 1 and 2 share the same DGM for ($\mathbf{W}=(W_1,W_2,W_3)$) and $A$: $W_1\sim N(0,1), W_2 \sim N(0.5, 1), W3 \sim Bernoulli(0.4)$; $P(A = 1 | \mathbf{W}) = \text{expit}(-0.48 + 0.96W_1 + 0.012W_2 + 1.08W_3$). For Scenario 1, $Y \sim N(\mu, 1),$ with $\mu = 10 + 2A + W1 + 0.7W_2 + 0.02W_3$; for Scenario 2,  $P(Y=1|A, \mathbf{W}) =  \mbox{expit}( -1.8 + 1.1A + 0.24W_1 + 0.08W_2 + 0.8W_3)$. The marginal ATE in Scenario 1 is  $\psi_{0,sim1}^{ATE}=2$. In Scenario 2, $\psi_{0,sim2}^{ATE}=0.2199, 0.2171, 0.2182,$ when $n=$ 100, 1000, 10,000, respectively. 

For Scenarios 1 and 2, we also ran simulations with treatment randomized 1:1 for $n=1000$, Scenarios 1a and 2a, respectively. Estimators again used the correct outcome model and  propensity score estimated from the true intercept-only regression model.  

\subsubsection{Scenario 3: Rare binary outcome}

Data with $n=10,000$ were generated as follows: 
$W_1 \sim N(0,1), W_2 \sim N(0.5, 1), W_3 \sim Bernoulli(0.4)$; $P(A =1 | \mathbf{W}) = \text{expit}(-0.72 - 0.72W_1 + 0.18W_2 + 0.81W_3)$, yielding approximately 43\% treated, with propensity score ranging from 0.025 to 0.975. $Y$ was generated from the logistic model $\text{logit}(P(Y=1|A, \mathbf{W}) = -4.9 -2A +  0.4W_1 - 4W_2 - 3W_3$, with a 5\% event rate. The marginal ATE  = -0.0247 and  marginal RR = 0.3454.  

\subsubsection{Scenario 4: Estimating a MSM regression coefficient}

We consider two MSM scenarios (Scenarios 4a and 4b) to estimate the logcOR, with the same DGM: $W_1 \sim N(0,1)$, $W_2 \sim N(0.5, 1)$, $W_3 \sim Bernoulli (0.4)$, $W_4$ is generated as $W_3 + N(0,1)$,  and $W_5$ is the binary indicator $W_1 > 0.2$ (approximately 42\% probability).  The correlation between $W_3$ and $W_4$ is $\rho_{W_3,W_4} = 0.44$, and between $W_1$ and $W_5$ is $\rho_{W_1, W_5} = 0.79$. Other variables are uncorrelated. $A$ is generated according to $\text{logit}(P(A=1|\mathbf{W})  = -0.4 + 0.8W_1 + .01W_2 + 0.9W_3)$, with propensity score range 0.34 -- 0.97 and mean 0.49.  The binary outcome (14\% event rate) was generated as $\text{logit}(P(Y=1|A,\mathbf{W})  = -2.5 + 1.1A + 0.24W_1 + .08 W_2 + 0.8W_3 - 0.3W_4 - 0.6W_5)$, equivalently expressed as $\text{logit}(P(Y=1|A,\mathbf{W})  = -2.5 + 1.1A + 0.24W_1 + .08 W_2 + 0.8W_3 - 0.3(W_3) - 0.3\epsilon - 0.6 I(W_4 < 0.2),$ with $\epsilon \sim N(0,1)$.  We let $\textbf{W}=(W_1,\ldots,W_5)$ denote the DGM full set of confounders, and $\mathbf{\widetilde{W}}$ denote the confounders in the MSM, which may or may not be the same as $\mathbf{W}$. We consider $n= 100, 1000, 10,000$.

In Scenario 4a, the MSM is the correct outcome  model, $\text{logit}(P(Y=1|A,\mathbf{W}) = \beta_0 + \beta_1A +  \beta_2W_1 + \beta_3W_2 + \beta_4W_3 + \beta_5W_4 + \beta_6W_5$. The MSM parameter values are the coefficients in the DGM model, $\mathbf{\beta} = (-2.5, 1.1, 0.24, 0.08, 0.8, -0.3, -0.6)$, with logcOR = 1.1.  

In Scenario 4b, the MSM conditions on all confounders but does not capture the true functional form of the outcome model. This MSM defines a projection of the underlying DGM onto a specified model, $E[Y|\mathbf{\widetilde{W}},A] =   \gamma_0 + \gamma_1A + \gamma_2W_1 + \gamma_3W_2 + \gamma_4W_3$. 
The MSM parameter values were established through simulation (see Supplemental Section \ref{sec:supp_synthetic}). The true value for logcOR is 1.084.

Datasets were created under each plasmode framework and analyzed by fitting a weighted logistic regression of $Y$ on the MSM using stabilized IPTW  with weights given by $wt_{stab, A=1} = p_A  A/P(A=1|W)$ for treated and $wt_{stab, A=0} = (1-p_A)(1-A) / [1-P(A=1|W)]$ for non-treated observations.  

\subsection{Results from synthetic datasets}

\subsubsection{Scenarios 1 and 2}

For Scenario 1 (Table \ref{tab:sim1-alln}), the \%bias and bias:SE for IPTW and glmPS are about 10-fold larger, or worse, for the Sample Treatment versus Generate Treatment approach. In several instances, the bias:SE remains larger than expected for a consistent method (i.e., rate $ \propto 1/\sqrt{n})$ for the Sample Treatment  but not  Generate Treatment approach. For the continuous outcome (Scenario 1), this is more apparent for IPTW than glmPS. For the binary outcome (Scenario 2),  the inflation of the bias:SE for IPTW and glmPS under the Sample Treatment approach appears smaller than for Scenario 1; however, the bias:SE does not decrease with sample size (Table \ref{tab:sim2-alln}). 

The magnitude of $\%$bias and bias:SE for the Match estimator is expected to be dominated by the quality of the matches, with increased continuity leading to larger expected bias; however, there were still notable differences in bias:SE between the two plasmode frameworks for the continuous outcome in Scenario 1. In Scenario 1, the bias:SE was again more than 10-fold worse and not clearly decreasing with sample size for the Sample Treatment approach. For Scenario 2, the performance of Match was more comparable between the two plasmode approaches.
 
 Under the two plasmode approaches, there are no theoretical reasons to expect a different numerical performance from glmCM or TMLE, whose consistency is leveraged from the correctly assumed outcome model, or from the unweighted Unadj estimator that does not adjust for confounding. For these three estimators, the \%bias and bias:SE are generally similar across $n$ for Scenarios 1 and 2. 




\subsubsection*{Coverage}
Wald-type 95\% confidence intervals (CIs) were calculated using the empirical SE. The bias of Unadj leads to poor coverage that tends towards 0 as $n$ increases (data not shown). For the continuous outcome, coverage of Match remained lower than the 95\% nominal level for all $n$ and IPTW estimators' coverage suffered at smaller sample sizes under the Sample Treatment approach; coverage remained at the nominal level for all estimators and $n$ with the Generate Treatment approach  (Figure S1a).  Coverage of all estimators was good under both plasmode sampling schemes for the binary outcome (Figure S1b).

\subsubsection*{Randomized treatment}

The Unadj estimator is known to be consistent in this RCT setting, yet under the Sample Treatment approach, it and the PS-based estimators (IPTW and glmPS) have a bias:SE larger than expected for a consistent estimator (Supplemental Tables~\ref{tab:sim1an1000},~\ref{tab:sim2an1000}). The impact is less pronounced when the outcome is binary. Under the Generate Treatment approach, all estimators' performance conforms to theoretical expectations.  The large bias:SE under the Sample Treatment approach manifests as poor coverage for the continuous outcome and slightly lower than expected coverage for the binary outcome  (Supplemental Figures S2, S3). 

\subsubsection{Scenario 3: Rare outcome results}

Increased outcome rarity increased the apparent problems for the Sample Treatment framework. There is a larger than expected bias:SE for the Sample Treatment approach for all PS-based estimators (Table~\ref{tab:sim3ATEn10000}).  The bias:SE for IPTW is nearly 200 times larger under the Sample Treatment versus Generate Treatment approach.  The bias:SE of glm-PS and Match is 8 times larger under the Sample Treatment approach.  The performance of Unadj, glmCM and TMLE have similar bias and SEs under the two plasmode approaches. 

\begin{figure}[htpb] 
   \centering
      \caption{Simulation Scenario 3 (rare binary outcome) Coverage for ATE} 
\includegraphics[width=6in]{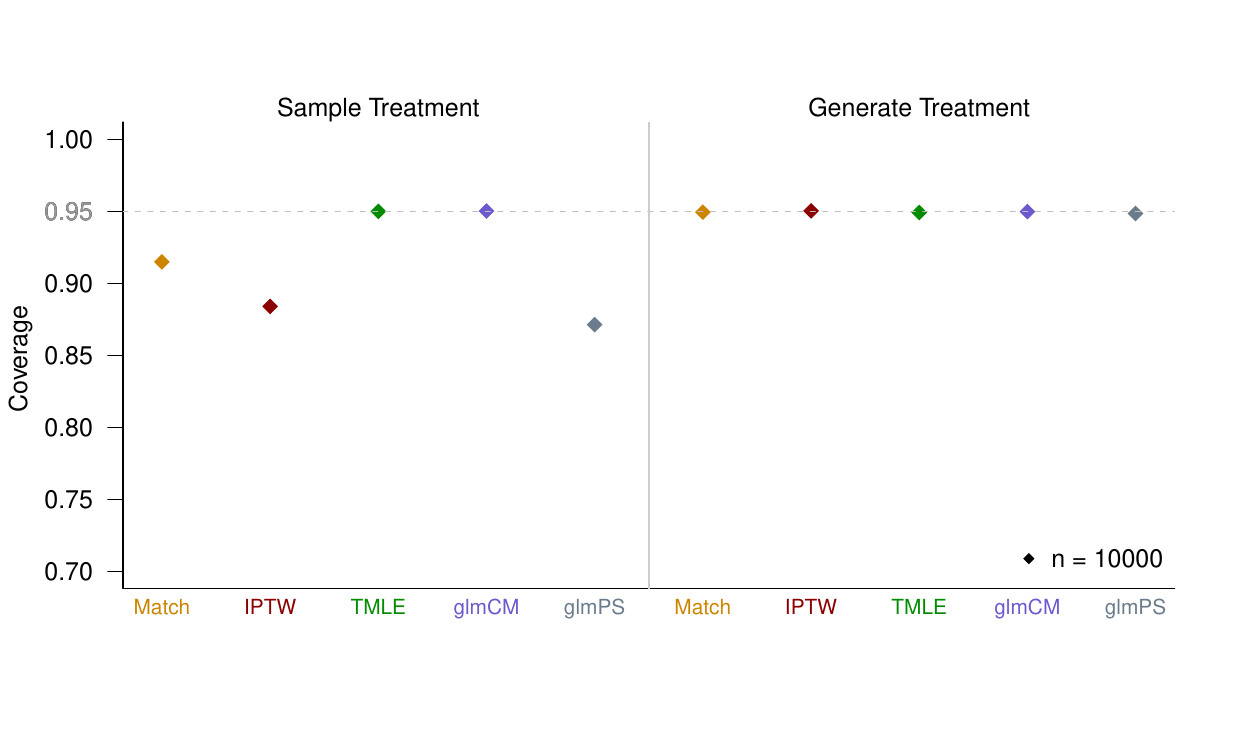} 
   \label{fig:sim3CovATE}
\end{figure}

\noindent A large bias:SE translated into lower than expected CI coverage for the PS-based estimators.  At this sample size all estimators have 100\% power to reject the null (data not shown), but for estimators where CI coverage is below 95\%, rejections are sometimes fueled by bias away from the null (Figure \ref{fig:sim3CovATE}). 


Table \ref{tab:sim3RRn10000} shows results for the RR. Results are identical for the estimator that uses the correct regression model (glmCM). For all other estimators except the TMLE, the bias:SE is smaller under the Generate Treatment approach.  For the TMLE, it is smaller under the Sample Treatment, but the same order of magnitude.   Coverage (Figure \ref{fig:sim3CovRR}) and power (data not shown) follow a similar pattern as for the ATE.

\subsubsection{Scenario 4: MSM results}
Scenario 4a results show that when the specified MSM is the true outcome model, both plasmode frameworks work correctly (Table~\ref{tab:sim4a-alln}).   

Scenario 4b results illustrate that performance of the plasmode approaches differ when the specified MSM is not equivalent to the true outcome model (Table~\ref{tab:sim4b-alln}).  The bias:SE under the Sample Treatment framework does not decrease at the expected rate with sample size, but under the Generate Treatment approach it does. 




\subsection{KPWA data simulation set-up}

Scenario 5 considers plasmode simulation using EHR data for 50,337 individuals 13 years and older initiating antidepressant medication or psychotherapy for treatment of depression at KPWA. These individuals have full information on the outcome of interest, self-harm, and key confounders. These data are a subset of data analyzed by \citet{williamson2024assessing} for a plasmode simulation to evaluate methods addressing confounder missingness. This FDA Sentinel Initiative project does not meet the criteria for human subject research as defined by Kaiser Permanente Washington Health Research Institute policies, Health and Human Services, and the FDA (\citeauthor{commonrule,OHRP}). The study involves public health surveillance activity defined by HHS regulation 45 CFR 46.102(l)(2).

The rate of 5-year self-harm or hospitalization is 10.3\%. Based on a logistic regression model fit to the data (Table~\ref{tab:original_coefficients}), the ATE comparing psychotherapy to antidepressant medication is -0.017 (\cite{williamson2024assessing}). We created two new plasmode DGMs with varying outcome rate and a larger treatment effect by modifying coefficients in these logistic regression models (Supplemental Materials Table~\ref{tab:original_coefficients}). For each DGM, we generated a new source dataset of size $n = 10,000$ (details in Supplementary Section~\ref{sec:supp_kpwa_analysis}). The ATEs in these two groups are -0.092 for the 15\% outcome  and -0.079 in the 5\% outcome dataset; the RRs are 0.512 (15\% outcome) and 0.062, respectively. 

We generated 100,000 datasets from these two source datasets under the Sample Treatment and Generate Treatment plasmode frameworks and evaluate the performance for the same estimators of the ATE and RR as described above.

\subsection{KPWA plasmode simulation results}
\begin{figure}[htpb]
    \centering
    
    \caption{Boxplots of the correlation between treatment and each covariate in the KPWA plasmode simulations. The outcome rate is given in the rows. The boxplots are over the 100,000 Monte Carlo replications. Color denotes the Sample Treatment (light gray) versus Generate Treatment (dark gray) plasmode approach. Abbreviations: Age sq.: age squared; ANX: anxiety disorder; Dx: diagnosis code; AUD: alcohol use disorder; Charlson = Charlson comorbidity index; PHQ8: Patient Health Questionnaire (PHQ) total score; PHQ9: PHQ ninth item; MH Hosp: hospitalization with mental health diagnosis code; SH: self-harm.}
    \includegraphics[width=1\linewidth]{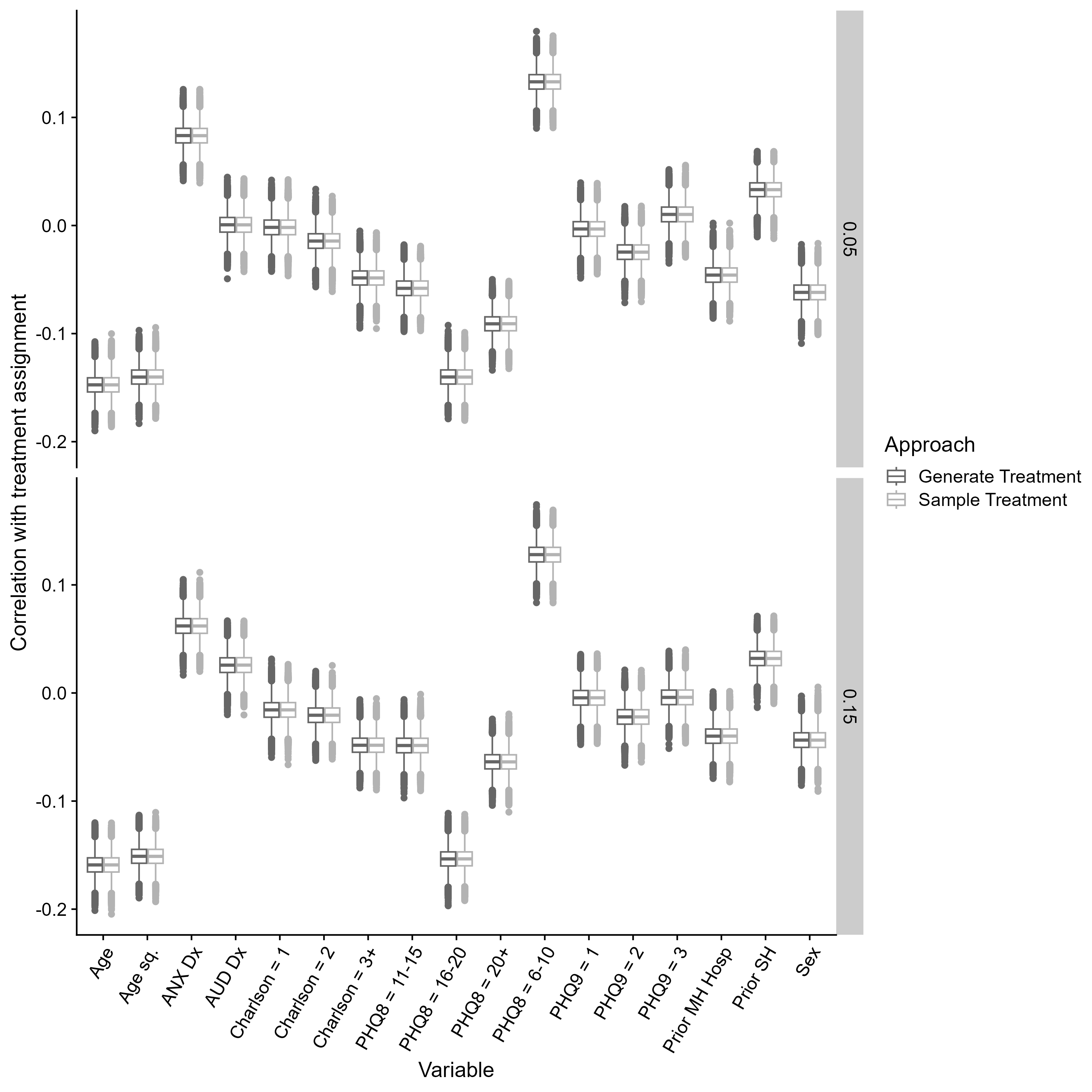}    
    \label{fig:kpwa_correlations}
\end{figure}

Results for the ATE and RR are in Tables \ref{tab:kpwa_plasmode_r0.05} and~\ref{tab:kpwa_plasmode_r0.15} for the 5\% and 15\% outcomes, respectively. Other estimands are in Tables~\ref{tab:kpwa_plasmode_r0.15_supp} and \ref{tab:kpwa_plasmode_r0.05_supp}. 
Results are largely similar to those for the purely synthetic data; however, there are two main differences. Firstly, TMLE has a larger bias:SE for the Sample Treatment than the Generate Treatment approach. While in the 15\% outcome, the bias:SE is of the expected order of magnitude, it is larger than expected under Sample Treatment for the 5\% outcome. Secondly, coverage is near the nominal level for all estimators besides the unadjusted estimator, regardless of plasmode strategy. These differences may be due to the complex regression models (including interactions with rare categorical variables), and the small number of events.  In Figure~\ref{fig:kpwa_correlations} we show the distribution of correlations between treatment and each covariate over Monte-Carlo replicates, which are nearly identical for the two plasmode approaches and indicate that sampling $W$ then generating $A$ does not affect the ability of plasmode simulations to capture correlation in the original data.





\section{Discussion}

When addressing causing questions about treatments using health care databases, analyses often need to address several complexities, including missing data, confounding, and potentially heterogeneous treatment effects.
Plasmode simulation is a valuable and practical tool to compare available analytical methods under realistic conditions in order to select the best performing method for the setting at hand; however, care must be taken to use an appropriate plasmode approach.   

We showed that the Sample Treatment plasmode framework leads to a biased evaluation of propensity-score based methods for a binary point treatment. This is because this sampling strategy creates an empirical distribution that does not adequately approximate the data generating distribution with respect to the treatment mechanism. We provided both theoretical justification and numerical evidence demonstrating this bias happens for the Sample Treatment but not Generate Treatment framework. This result readily generalizes to other inverse-probability of censoring and treatment weighted estimators.

Simulation studies were generally performed under correct model specifications to emphasize that the observed discrepancies were driven by underlying deficiencies in the plasmode framework, not by complexities in the data-generating process or shortcomings of a particular analytical procedure.  We showed the Generate Treatment framework had good numerical performance across all scenarios, with the demonstrated expected asymptotic regularity for all estimators under study. Differences between the two plasmode frameworks were more notable for continuous versus binary outcomes and for more rare binary outcomes.  

Finally, we note that plasmode datasets that preserve the underlying structure of real-world data are most relevant for characterizing estimator performance.  We recommend avoiding simplistic parametric models for the DGM by flexible modeling or machine learning. These models can be used to generate synthetic treatments and outcomes after re-sampling covariates. We showed in a plasmode simulation based on real-world data that the Generate Treatment approach preserved complexity present in the original data. Plasmode simulation studies conducted in this manner will provide a reliable basis for evaluating and comparing estimator performance.

\section*{Funding}
The author(s) disclosed receipt of the following financial support for the research, authorship, and/or publication of this article: This project was supported by Task Order 75F40123F19006 under Master Agreement 75F40119D10037 from the US Food and Drug Administration (FDA) and  the U.S. National Institutes of Health (NIH) grant R37-AI131771.

\section*{Acknowledgments}
The contents are those of the authors and do not necessarily represent the official views of, nor endorsement, by FDA/HHS, National Institutes of Health, or the U.S. Government.

\section*{Declaration of competing interest} None.


\section*{Data availability}
The datasets generated and analyzed during this study are not publicly available because they contain detailed information from the electronic health records in the health systems participating in this study and are governed by HIPAA. Data are however available from the authors upon reasonable request, with permission of all health systems involved and fully executed data use agreement.


\newpage

\begin{table}[htbp]
   \centering
      \caption{Data generating mechanisms for plasmode simulation approaches.}
   \begin{tabular}{@{} lll @{}} 
      \toprule
	&\bf{Sample Treatment}& 	\bf{Generate Treatment} \\
	\midrule
Covariates & Sample $W$ with replacement& Sample $W$ with replacement\\
Treatment &	Sample $A=a$ along with $W$ & Generate $A^{\#} \sim f_A(W, U_A)$\\
Outcome &	Generate $Y^{\#} \sim f_Y(A,W, U_Y)$  & Generate  $Y^{\#} \sim f_Y(A^{\#},W, U_Y)$ \\
      \bottomrule
   \end{tabular}
   \label{tab:resampling}
\end{table}

 \clearpage

 \begin{table}[!htbp]
\caption{Simulation Scenario 1: Estimate ATE for continuous outcome}
\label{tab:sim1-alln}
\begin{center}
\begin{tabular}{lrrrrcrrrr}
\hline\hline
\multicolumn{1}{l}{\bfseries }&\multicolumn{4}{c}{\bfseries Sample Treatment}&\multicolumn{1}{c}{\bfseries }&\multicolumn{4}{c}{\bfseries Generate Treatment}\tabularnewline
\cline{2-5} \cline{7-10}
\multicolumn{1}{l}{}&\multicolumn{1}{c}{\% Bias}&\multicolumn{1}{c}{SE}&\multicolumn{1}{c}{RMSE}&\multicolumn{1}{c}{Bias:SE}&\multicolumn{1}{c}{}&\multicolumn{1}{c}{\% Bias}&\multicolumn{1}{c}{SE}&\multicolumn{1}{c}{RMSE}&\multicolumn{1}{c}{Bias:SE}\tabularnewline
\hline
{\bfseries $n=100$}&&&&&&&&&\tabularnewline
~~Unadj&$159.29$&$1.337$&$3.455$&$2.382$&&$159.60$&$1.344$&$3.463$&$2.376$\tabularnewline
~~Match&$-10.78$&$0.818$&$0.846$&$0.264$&&$  3.80$&$0.843$&$0.846$&$0.090$\tabularnewline
~~IPTW&$-19.65$&$0.612$&$0.727$&$0.642$&&$  2.08$&$0.478$&$0.479$&$0.087$\tabularnewline
~~TMLE&$ -0.15$&$0.236$&$0.236$&$0.013$&&$ -0.15$&$0.234$&$0.234$&$0.013$\tabularnewline
~~glmCM&$ -0.01$&$0.224$&$0.224$&$0.001$&&$  0.01$&$0.223$&$0.223$&$0.001$\tabularnewline
~~glmPS&$ -2.41$&$0.248$&$0.252$&$0.195$&&$  0.10$&$0.236$&$0.236$&$0.009$\tabularnewline
\hline
{\bfseries $n=1000$}&&&&&&&&&\tabularnewline
~~Unadj&$ 19.53$&$0.470$&$0.611$&$0.831$&&$ 19.33$&$0.469$&$0.608$&$0.824$\tabularnewline
~~Match&$-18.01$&$0.548$&$0.655$&$0.658$&&$  0.85$&$0.419$&$0.419$&$0.041$\tabularnewline
~~IPTW&$ -7.28$&$0.235$&$0.276$&$0.620$&&$  0.18$&$0.221$&$0.221$&$0.016$\tabularnewline
~~TMLE&$ -0.13$&$0.076$&$0.076$&$0.034$&&$ -0.06$&$0.076$&$0.076$&$0.016$\tabularnewline
~~glmCM&$  0.00$&$0.071$&$0.071$&$0.001$&&$ -0.01$&$0.071$&$0.071$&$0.002$\tabularnewline
~~glmPS&$ -0.08$&$0.072$&$0.072$&$0.021$&&$  0.00$&$0.071$&$0.071$&$0.001$\tabularnewline
\hline
{\bfseries $n=10000$}&&&&&&&&&\tabularnewline
~~Unadj&$ 43.62$&$0.141$&$0.884$&$6.168$&&$ 43.57$&$0.142$&$0.883$&$6.138$\tabularnewline
~~Match&$  3.43$&$0.154$&$0.169$&$0.445$&&$  0.00$&$0.117$&$0.117$&$0.001$\tabularnewline
~~IPTW&$  0.09$&$0.056$&$0.056$&$0.033$&&$  0.01$&$0.058$&$0.058$&$0.003$\tabularnewline
~~TMLE&$  0.00$&$0.023$&$0.023$&$0.003$&&$  0.00$&$0.024$&$0.024$&$0.004$\tabularnewline
~~glmCM&$  0.00$&$0.022$&$0.022$&$0.002$&&$  0.00$&$0.022$&$0.022$&$0.001$\tabularnewline
~~glmPS&$ -0.01$&$0.022$&$0.022$&$0.009$&&$  0.00$&$0.022$&$0.022$&$0.001$\tabularnewline
\hline
\multicolumn{10}{l}{\footnotesize ATE: average treatment effect; SE: Standard Error; RMSE: root mean squared error; Unadj: unadjusted} \tabularnewline
\multicolumn{10}{l}{\footnotesize IPTW: inverse probability of treatment weighted; TMLE: targeted maximum likelihood estimation;}\tabularnewline
\multicolumn{10}{l} {\footnotesize glmCM: generalized linear model correct model; glmPS: generalized linear model propensity score} \tabularnewline
\end{tabular}\end{center}
\end{table}

 \clearpage

  \begin{table}[!htbp]
\caption{Simulation Scenario 2: Estimate ATE for binary outcome}
\label{tab:sim2-alln} 
\begin{center}
\begin{tabular}{lrrrrcrrrr}
\hline\hline
\multicolumn{1}{l}{\bfseries }&\multicolumn{4}{c}{\bfseries Sample Treatment}&\multicolumn{1}{c}{\bfseries }&\multicolumn{4}{c}{\bfseries Generate Treatment}\tabularnewline
\cline{2-5} \cline{7-10}
\multicolumn{1}{l}{}&\multicolumn{1}{c}{\% Bias}&\multicolumn{1}{c}{SE}&\multicolumn{1}{c}{RMSE}&\multicolumn{1}{c}{Bias:SE}&\multicolumn{1}{c}{}&\multicolumn{1}{c}{\% Bias}&\multicolumn{1}{c}{SE}&\multicolumn{1}{c}{RMSE}&\multicolumn{1}{c}{Bias:SE}\tabularnewline
\hline
{\bfseries $n=100$}&&&&&&&&&\tabularnewline
~~Unadj&$29.25$&$0.091$&$0.111$&$0.708$&&$29.71$&$0.091$&$0.112$&$0.720$\tabularnewline
~~Match&$ 1.15$&$0.129$&$0.129$&$0.020$&&$ 1.11$&$0.119$&$0.119$&$0.020$\tabularnewline
~~IPTW&$-2.23$&$0.110$&$0.110$&$0.044$&&$ 0.51$&$0.106$&$0.106$&$0.011$\tabularnewline
~~TMLE&$ 0.11$&$0.106$&$0.106$&$0.002$&&$ 0.16$&$0.106$&$0.106$&$0.003$\tabularnewline
~~glmCM&$ 0.10$&$0.101$&$0.101$&$0.002$&&$ 0.18$&$0.101$&$0.101$&$0.004$\tabularnewline
~~glmPS&$-0.34$&$0.101$&$0.101$&$0.007$&&$-0.02$&$0.101$&$0.101$&$0.000$\tabularnewline
\hline
{\bfseries $n=1000$}&&&&&&&&&\tabularnewline
~~Unadj&$32.81$&$0.028$&$0.077$&$2.538$&&$33.05$&$0.028$&$0.077$&$2.556$\tabularnewline
~~Match&$ 0.13$&$0.043$&$0.043$&$0.006$&&$ 0.27$&$0.039$&$0.039$&$0.015$\tabularnewline
~~IPTW&$-0.66$&$0.034$&$0.034$&$0.042$&&$ 0.07$&$0.034$&$0.034$&$0.005$\tabularnewline
~~TMLE&$-0.06$&$0.034$&$0.034$&$0.004$&&$ 0.00$&$0.034$&$0.034$&$0.000$\tabularnewline
~~glmCM&$-0.05$&$0.032$&$0.032$&$0.004$&&$ 0.00$&$0.032$&$0.032$&$0.000$\tabularnewline
~~glmPS&$-0.15$&$0.032$&$0.032$&$0.010$&&$-0.05$&$0.032$&$0.032$&$0.003$\tabularnewline
\hline
{\bfseries $n=10000$}&&&&&&&&&\tabularnewline
~~Unadj&$32.18$&$0.009$&$0.071$&$7.909$&&$32.22$&$0.009$&$0.071$&$7.862$\tabularnewline
~~Match&$ 0.29$&$0.013$&$0.013$&$0.048$&&$ 0.03$&$0.012$&$0.012$&$0.006$\tabularnewline
~~IPTW&$ 0.34$&$0.010$&$0.010$&$0.071$&&$ 0.02$&$0.011$&$0.011$&$0.005$\tabularnewline
~~TMLE&$ 0.02$&$0.010$&$0.010$&$0.003$&&$ 0.02$&$0.011$&$0.011$&$0.004$\tabularnewline
~~glmCM&$ 0.01$&$0.010$&$0.010$&$0.001$&&$ 0.01$&$0.010$&$0.010$&$0.003$\tabularnewline
~~glmPS&$-0.03$&$0.010$&$0.010$&$0.006$&&$ 0.00$&$0.010$&$0.010$&$0.001$\tabularnewline
\hline
\multicolumn{10}{l}{\footnotesize ATE: average treatment effect; SE: Standard Error; RMSE: root mean squared error; Unadj: unadjusted} \tabularnewline
\multicolumn{10}{l}{\footnotesize IPTW: inverse probability of treatment weighted; TMLE: targeted maximum likelihood estimation;}\tabularnewline
\multicolumn{10}{l} {\footnotesize glmCM: generalized linear model correct model; glmPS: generalized linear model propensity score} \tabularnewline
\end{tabular}\end{center}
\end{table}

   \clearpage

 \begin{table}[!htbp]
\caption{Simulation Scenario 3: Estimate ATE for rare binary outcome, $n=10000$\label{tab:sim3ATEn10000}} 
\begin{center}
\begin{tabular}{lrrrrcrrrr}
\hline\hline
\multicolumn{1}{l}{\bfseries }&\multicolumn{4}{c}{\bfseries Sample Treatment}&\multicolumn{1}{c}{\bfseries }&\multicolumn{4}{c}{\bfseries Generate Treatment}\tabularnewline
\cline{2-5} \cline{7-10}
\multicolumn{1}{l}{}&\multicolumn{1}{c}{\% Bias}&\multicolumn{1}{c}{SE}&\multicolumn{1}{c}{RMSE}&\multicolumn{1}{c}{Bias:SE}&\multicolumn{1}{c}{}&\multicolumn{1}{c}{\% Bias}&\multicolumn{1}{c}{SE}&\multicolumn{1}{c}{RMSE}&\multicolumn{1}{c}{Bias:SE}\tabularnewline
\hline
Unadj&$41.698$&$0.003$&$0.011$&$3.369$&&$34.918$&$0.003$&$0.009$&$2.793$\tabularnewline
Match&$ 8.493$&$0.004$&$0.004$&$0.555$&&$-0.103$&$0.003$&$0.003$&$0.007$\tabularnewline
IPTW&$ 9.362$&$0.003$&$0.004$&$0.747$&&$-0.052$&$0.003$&$0.003$&$0.004$\tabularnewline
TMLE&$-0.001$&$0.003$&$0.003$&$0.000$&&$-0.052$&$0.003$&$0.003$&$0.005$\tabularnewline
glmCM&$-0.044$&$0.002$&$0.002$&$0.004$&&$-0.033$&$0.002$&$0.002$&$0.003$\tabularnewline
glmPS&$ 9.939$&$0.003$&$0.004$&$0.813$&&$ 1.343$&$0.003$&$0.003$&$0.108$\tabularnewline
\hline
\multicolumn{10}{l}{\footnotesize ATE: average treatment effect; SE: Standard Error; RMSE: root mean squared error; Unadj: unadjusted;} \tabularnewline
\multicolumn{10}{l}{\footnotesize IPTW: inverse probability of treatment weighted; TMLE: targeted maximum likelihood estimation;}\tabularnewline
\multicolumn{10}{l} {\footnotesize glmCM: generalized linear model correct model; glmPS: generalized linear model propensity score}\tabularnewline
\end{tabular}\end{center}
\end{table}

   \clearpage

 \begin{table}[!htbp]
\caption{Simulation Scenario 4b: Estimate logcOR when MSM is not equivalent to true outcome regression\label{tab:sim4b-alln}} 
\begin{center}
\begin{tabular}{lrrrrcrrrr}
\hline\hline
\multicolumn{1}{l}{\bfseries }&\multicolumn{4}{c}{\bfseries Sample Treatment}&\multicolumn{1}{c}{\bfseries }&\multicolumn{4}{c}{\bfseries Generate Treatment}\tabularnewline
\cline{2-5} \cline{7-10}
\multicolumn{1}{l}{}&\multicolumn{1}{c}{\% Bias}&\multicolumn{1}{c}{SE}&\multicolumn{1}{c}{RMSE}&\multicolumn{1}{c}{Bias:SE}&\multicolumn{1}{c}{}&\multicolumn{1}{c}{\% Bias}&\multicolumn{1}{c}{SE}&\multicolumn{1}{c}{RMSE}&\multicolumn{1}{c}{Bias:SE}\tabularnewline
\hline
$n=100$&$60.424$&$2.829$&$2.904$&$0.232$&&$50.114$&$2.712$&$2.766$&$0.200$\tabularnewline
$n=1000$&$ 4.189$&$0.228$&$0.232$&$0.199$&&$ 1.323$&$0.229$&$0.229$&$0.063$\tabularnewline
$n=10000$&$ 0.780$&$0.071$&$0.072$&$0.118$&&$ 0.104$&$0.071$&$0.071$&$0.016$\tabularnewline
\hline
\multicolumn{10}{l}{\footnotesize logcOR: log conditional odds ratio; MSM: marginal structural model; SE: Standard Error; RMSE: root} \tabularnewline
\multicolumn{10}{l}{\footnotesize mean squared error; Unadj: unadjusted; IPTW: inverse probability of treatment weighted; TMLE: targeted}\tabularnewline
\multicolumn{10}{l} {\footnotesize maximum likelihood estimation; glmCM: generalized linear model correct model; glmPS: generalized linear}\tabularnewline
\multicolumn{10}{l}{\footnotesize model propensity score}
\end{tabular}\end{center}
\end{table}

   \clearpage

 \begin{table}[!htbp]
\centering
\caption{Results of the KPWA-based plasmode simulations for the 5\% outcome (ATE and RR).\label{tab:kpwa_plasmode_r0.05}}
\centering
\resizebox{\ifdim\width>\linewidth\linewidth\else\width\fi}{!}{
\fontsize{11}{13}\selectfont
\begin{threeparttable}
\begin{tabular}[t]{llllllrllllr}
\toprule
\multicolumn{2}{c}{ } & \multicolumn{5}{c}{Sample Treatment} & \multicolumn{5}{c}{Generate Treatment} \\
\cmidrule(l{3pt}r{3pt}){3-7} \cmidrule(l{3pt}r{3pt}){8-12}
Estimand & Estimator & \% Bias & SE & RMSE & bias:SE & CP & \% Bias & SE & RMSE & bias:SE & CP\\
\midrule
 & Unadj & 10.964 & 0.004 & 0.010 & 2.130 & 43.4 & 11.191 & 0.004 & 0.010 & 2.169 & 41.8\\
\cmidrule{2-12}
 & Match & 0.403 & 0.005 & 0.005 & 0.064 & 94.9 & -0.245 & 0.005 & 0.005 & 0.042 & 95.0\\
\cmidrule{2-12}
 & IPTW & 1.189 & 0.005 & 0.005 & 0.195 & 95.3 & -0.219 & 0.004 & 0.004 & 0.042 & 95.1\\
\cmidrule{2-12}
 & TMLE & 0.571 & 0.005 & 0.005 & 0.096 & 95.1 & 0.012 & 0.004 & 0.004 & 0.002 & 95.1\\
\cmidrule{2-12}
 & glmCM & -0.175 & 0.004 & 0.004 & 0.034 & 95.3 & -0.182 & 0.004 & 0.004 & 0.036 & 95.1\\
\cmidrule{2-12}
\multirow{-6}{*}{\raggedright\arraybackslash ATE} & glmPS & -2.553 & 0.004 & 0.004 & 0.519 & 91.9 & -2.874 & 0.004 & 0.004 & 0.586 & 90.9\\
\cmidrule{1-12}
 & Unadj & -20.563 & 0.011 & 0.017 & 1.166 & 77.9 & -20.875 & 0.011 & 0.017 & 1.188 & 77.3\\
\cmidrule{2-12}
 & Match & -3.705 & 0.019 & 0.019 & 0.123 & 95.6 & -1.548 & 0.018 & 0.018 & 0.054 & 95.5\\
\cmidrule{2-12}
 & IPTW & 0.328 & 0.016 & 0.016 & 0.013 & 95.2 & 0.340 & 0.016 & 0.016 & 0.014 & 95.2\\
\cmidrule{2-12}
 & TMLE & -0.555 & 0.016 & 0.016 & 0.022 & 95.2 & 0.062 & 0.016 & 0.016 & 0.002 & 95.1\\
\cmidrule{2-12}
 & glmCM & 0.362 & 0.014 & 0.014 & 0.016 & 95.1 & 0.326 & 0.014 & 0.014 & 0.014 & 95.1\\
\cmidrule{2-12}
\multirow{-6}{*}{\raggedright\arraybackslash RR} & glmPS & 4.675 & 0.014 & 0.015 & 0.201 & 94.6 & 5.558 & 0.015 & 0.015 & 0.237 & 94.3\\
\bottomrule
\end{tabular}
\begin{tablenotes}
\item Abbreviations: KPWA: Kaiser Permanente Washington; ATE: average treatment effect; RR: relative risk; SE: standard error; RMSE: root mean squared error; Bias:SE: ratio of bias to standard error; CP: coverage probability; Unadj: unadjusted; Match: propensity score matching; IPTW: inverse probability of treatment weighting; TMLE: targeted maximum likelihood estimation; glmCM: generalized linear model, correctly specified; glmPS: generalized linear model, adjusted for propensity score. Results for n = 10,000 ($1/\sqrt{n}=$0.01) with 100,000 Monte-Carlo iterations. The true value of the ATE is -0.079; the true value of the RR is 0.062.
\end{tablenotes}
\end{threeparttable}}
\end{table}

   \clearpage

   \clearpage

\clearpage
\newpage

\ifarXiv
\renewcommand{\thesection}{S\arabic{section}}
\renewcommand{\thetable}{S\arabic{table}}  
\renewcommand{\thefigure}{S\arabic{figure}}
\renewcommand{\figurename}{Supplemental Figure} 
\renewcommand{\tablename}{Supplemental Table} 
\renewcommand{\theequation}{S\arabic{equation}}
\setcounter{table}{0}
\setcounter{figure}{0}
\setcounter{tocdepth}{2}
\setcounter{section}{0}
\setcounter{equation}{0}


\newif\ifnotarXiv
\notarXivfalse

\ifarXiv
\section*{SUPPLEMENTARY MATERIALS}
\fi


\section{Mathematical details for the manuscript Section 4:  Theoretical problems for the Sample Treatment plasmode framework} \label{sec:theory}
\subsection{Overview}
Consider the setting discussed in this article in which we observe a sample of $n$ i.i.d.~observations $O_i=(W_i,A_i,Y_i)\sim_{iid} P_0$, where $W$ is a covariate vector, $A$ is a treatment indicator, and $Y$ is the outcome. Consider a statistical target estimand given by $\Psi(P_0)=E_{P_0} E_{P_0}(Y\mid A=1,W)$, which can be causally interpreted as the marginal mean outcome under treatment $A=1$, $EY_1$, under causal assumptions. We focus on the simple IPTW estimator $\hat{\Psi}(P_n)=1/n\sum_{i=1}^n A_iY_i/\hat{g}(P_n)$ for some estimator $\hat{g}(P_n)$ of the propensity score $\bar{g}_0(W)=P_0(A=1\mid W)$.
For simplicity, let $\hat{g}(P_n)$ be a maximum likelihood estimator based on a parametric  logistic regression model for $\bar{g}_0(W)$ and suppose that the parametric model is correctly specified. Then, we know that this IPTW estimator is consistent and asymptotically linear with a known influence curve. In particular, $Z_n=n^{1/2}(\hat{\Psi}(P_n)-\Psi(P_0))$ converges to a mean zero normal distribution $Z_0\sim N(0,\sigma^2_0)$ for a specified variance $\sigma^2_0$.
Thus, the IPTW estimator has negligible bias converging to zero at a faster rate than $n^{-1/2}$ (even at rate $n^{-1}$ here), and has a well understood normal limit distribution that can be used for construction of valid Wald type confidence intervals. 

Consider now a plasmode simulation study conducted using the \emph{Sample Treatment} approach described in the main article, which one might also call a model based bootstrap. In this study we use empirical sampling of $(W^{\#},A^{\#})$ from the empirical distribution of $(W_i,A_i)$, $i=1,\ldots,n$ then, given $(W^{\#},A^{\#})$,  simulate $Y^{\#}$ from a particular conditional distribution ${\bf p}_{Y,n}$  of $Y$, given $(W,A)$. Let's denote this random variable with $O^{\#}=(W^{\#},A^{\#},Y^{\#})$ and its data distribution with ${\bf P}_n$. Note that the density of ${\bf P}_n$ is given by ${\bf p}_n(w,a,y)=p_n(w,a){\bf p}_{Y,n}(y\mid w,a)$. Thus, $(W^{\#},A^{\#})$ has a  discrete distribution that puts mass $1/n$ on each $(W_i,A_i)$ in the original sample  from $P_0$, while ${\bf p}_{Y,n}$ might be a consistent estimator of the true $p_{Y,0}$ obtained from the original sample.  Let $P_n^{\#}$ be the empirical distribution of a random  sample $O_i^{\#}$, $i=1,\ldots,n$.
Under the distribution ${\bf P}_n$, the true parameter value might now be defined by ${\bf \psi}_n\equiv 1/n \sum_{i=1}^n \bar{Q}_n(W_i)$ where $\bar{Q}_n(w)$ is the conditional mean of $Y$, given $W=w,A=1$, under ${\bf p}_{Y,n}$.
 One might now evaluate the sampling distribution of the standardized IPTW estimator $Z_n^{\#}\equiv n^{1/2}(\hat{\Psi}(P_n^{\#})-{\bf \psi}_n)$, based on this sample of $n$ draws from ${\bf P}_n$, as a measure of performance of the IPTW estimator in this simulation. The natural assumption one would like to make  is that the behavior of this estimator under sampling from ${\bf P}_n$ resembles the behavior of the estimator under sampling from $P_0$ as ${\bf p}_{Y,n}$ approximates $p_{Y,0}$ the true conditional distribution of $Y$, given $W,A$. Formally, given $P_n$, one would assume that the sampling distribution of $Z_n^{\#}$ approximates the sampling distribution of $Z_n$ that is known to converge to $Z_0$.  The latter is the formal way of stating that the chosen bootstrap distribution consistently estimates the sampling distribution of our estimator. But is this actually true? 

Let's first make clear what will be true. We will have that, conditional on $P_n$, 
$n^{1/2}(\hat{\Psi}(P_n^{\#})-\hat{\Psi}({\bf P}_n))$ converges in distribution to the same normal limit distribution as  $n^{1/2}(\hat{\Psi}(P_n)-\Psi(P_0))$, under the assumption that ${\bf p}_{Y,n}$ consistently estimates $p_{Y,0}$. This is nothing else than stating that the model-based bootstrap consistently estimates the sampling distribution of the IPTW estimator. This is even true under sampling from $P_n$, in which case it just states the   so called nonparametric boostrap consistently estimates the sampling distribution of the IPTW estimator. The regularity conditions for consistency of a model based bootstrap like sampling from ${\bf P}_n$ are even weaker than that of the nonparametric bootstrap (basically relying on so called compact differentiability of the estimator as a functional of the empirical measure $P_n$ \cite{Gill89, vanderVaart&Wellner96}).
From this result we already can conclude that \[
Z_n^{\#}=n^{1/2}(\hat{\Psi}(P_n^{\#})-\Psi({\bf P}_n))=n^{1/2}(\hat{\Psi}(P_n^{\#})-\hat{\Psi}({\bf P}_n))+n^{1/2}B_n,\]
 where the first term converges to the desired normal limit distribution and, conditional on $P_n$,  $B_n\equiv (\hat{\Psi}({\bf P}_n)-{\bf \psi}_n)$ is a bias term. So our $Z_n^{\#}$ will only consistently estimate the normal distribution sampling distribution if $n^{1/2}B_n\rightarrow_p 0$. However, that is not true. Relatively straightforward algebra presented below shows that $n^{1/2}B_n=n^{1/2}(P_n-P_0)f+o_P(1)$ behaves as  a standardized empirical mean of a function $f(O)$. To keep it simple and as demonstration of the latter statement, let's say that $g_0$ is known and $\hat{g}(P_n)=g_0$ so that the IPTW estimator of the marginal mean outcome under treatment is just an empirical mean of $Y_iA_i/\bar{g}_0(W_i)$. 
 In this perfect scenario, we have (exact)
 \[
 n^{1/2}B_n=
 n^{-1/2}\sum_{i=1}^n \bar{Q}_n(W_i)/\bar{g}_0(W_i)(A_i-\bar{g}_0(W_i)).\]
 Conditional on $P_n$, this is a sum of independent identically distributed random variables so that it converges to a normal mean zero distribution. 
 So, given $P_n$, $n^{1/2}B_n$ is a realization of a draw from an approximately normally distributed random variable. Clearly, this is not going to zero even as sample size tends to infinity, which means that the sampling distribution $Z_n^{\#}$ reflects a non-negligible bias $B_n$. In particular, 95\% confidence intervals based on $P_n^{\#}$ that ignore this bias will not contain the true ${\bf \psi}_n$ with probability 0.95, not even as sample size approximates infinity.
 
So what happened?  The problem was that our true data distribution ${\bf P}_n$ is such that
$\hat{\Psi}({\bf P}_n)\not =\Psi({\bf P}_n)$. That is, our estimator applied to an infinite sample from ${\bf P}_n$ is still not equal to our true parameter value. 
This is a clear warning that  one is simulating from a distribution under which our target parameter is ill defined. That is, $E_{{\bf P}_n}(Y\mid W,A=1)$ does not exist due to the fact that for various $W_i$ we have that ${\bf P}_n(W^{\#}=W_i,A_i^{\#}=1)$ equals zero. We then assign a definition that superficially makes sense, namely $P_n \bar{Q}_n$, but that is not dealing with the fundamental issue. The fact that this issue disappears in some sense asymptotically also does  not solve the problem since a bias of size $n^{-1/2}$ remains, thereby affecting MSE and coverage. The same argument applies to the IPTW estimator of the marginal mean outcome under the comparator treatment, $EY_0$, and any causal contrast such as the ATE, RR, etc.

Below we formally show that this bias for the Sample Treatment plasmode framework occurs for IPTW estimators and also for propensity score matching estimators. In fact, it also causes a bias in any double robust estimator such as an A-IPTW or TMLE when the outcome regression is inconsistently estimated. We also show that the Generate Treatment plasmode framework will be a consistent procedure. Our simulations show that even  for estimators that solely rely on the outcome regression estimation, the finite sample bias  under Sample Treatment simulations might be meaningfully larger than under Generate Treatment simulations.

\subsection{Formal mathematical argument}

\subsection*{Preliminaries}

Consider a sample of $n$ i.i.d. copies $O_1,\ldots,O_n$ of $O=(W,A,Y)\sim P_0$ and let $P_n$ be the empirical measure of $O_1,\ldots,O_n$. Let $A\in \{0,1\}$ be binary; $g_0(1\mid W)=P_0(A=1\mid W)$ and $\bar{Q}_0(W)=E_0(Y\mid A=1,W)$. We will also use short-hand notation $g_0(W)$ for $g_0(1\mid W)$.
Suppose the target parameter of interest $\Psi(P_0)=E_0\bar{Q}_0(W)$. We assume the strong positivity assumption $g_0(1\mid W)>\delta>0$, $P_{W,0}$-a.e., for some $\delta>0$ and let ${\cal G}$ be a parametric model for $g_0$ so that it is known that $g_0\in {\cal G}$. This defines a statistical model ${\cal M}({\cal G})=\{p_Wg_Ap_Y: p_W,p_Y,g_A\in {\cal G}\}$  for $P_0$ that only assumes the parametric model on $g_0$ while it leaves the distribution of the covariates and the conditional distribution $P_{Y,0}$ unspecified. We also define a nonparametric model ${\cal G}(CAR)$ for $g_0$ that makes no assumptions on $g_0$: this model corresponds with only assuming CAR on the missingness mechanism when we define the full-data $X=(W,Y(0),Y(1))$ and missingness variable is given by $A$, so that $O=(W,A,Y(A))$ and CAR states $P(A=1\mid X)=P(A=1|W)$ \citep{vanderlaan&robins03}.
 
 \subsection*{The IPTW estimator and its asymptotics}
Let $\hat{g}(P_n)$ be  maximum likelihood estimator of $g_0$ according to the correctly specified parametric model ${\cal G}$ with tangent space $T_{\cal G}(P_0)$. For example, if we use a logistic regression model ${\cal G}$ defined by $\mbox{Logit}P(A=1\mid W)=\beta^{\top}\phi(W)$ for a vector $\phi(W)=(\phi_j(W):j=1,\ldots,J)$ of covariates, then this tangent space $T_{{\cal G}}(P_0)=\{\{\sum_j\alpha(j)\phi_j(W)\}(A-g_0(1\mid W)):\alpha\in \openr^J\}$ is the linear span of the standard scores $\phi_j(W)(A-g_0(1\mid W))$ of the coefficients $\beta_j$ in the logistic regression model. By properties of parametric maximum likelihood estimators for correct models, we have that $\hat{g}(P_n)$ is $n^{-1/2}$-consistent: $\pl \hat{g}(P_n)-g_0\pl_{\infty}=O_P(n^{-1/2})$. 

A possible root-$n$-consistent estimator of $\psi_0$ is given by the IPTW estimator $\psi_n=\hat{\Psi}(P_n)=E_{P_n} AY/\hat{g}(P_n)$. This estimator is asymptotically linear with influence curve $D_{IPTW,P_0}=D_{P_0}-D_{{\cal G},P_0}$, where $D_{P_0}(O)=AY/g_0(W)-\psi_0$ and $D_{{\cal G},P_0}=\Pi(D_{P_0}\mid T_{{\cal G}}(P_0))$ denotes the projection of $D_{P_0}$ onto sub-Hilbert space $T_{{\cal G}}(P_0)\subset L^2_0(P_0)$ of the Hilbert space $L^2_0(P_0)$ with inner product $\langle h_1,h_2\rangle_{P_0}P_0 h_1h_2$ the covariance operator under $P_0$. This is due to a well known result that estimating the orthogonal nuisance parameter $g_0$ with a maximum likelihood procedure $\hat{g}(P_n)$ in an estimator such as an IPTW estimator results in a subtraction $D_{{\cal G},P_0}$ from the influence curve of the estimator when using the true $g_0$ (Theorem 2.3, \cite{vanderlaan&robins03}).

In particular, the IPTW estimator gains in efficiency by  being an MLE for a larger and larger model ${\cal G}$, and, if ${\cal G}={\cal G}(CAR)$ is the nonparametric model then the IPTW estimator is asymptotically linear with influence curve equal to the canonical gradient $D^*_{P_0}=D_{P_0}-D_{CAR,P_0}$ of the pathwise derivative of  $\Psi:{\cal M}({\cal G})\rightarrow\openr$ at $P_0$. Thus, in that case the IPTW estimator is known to be asymptotically efficient. For continuous $W$, such an NPMLE $\hat{g}$ would not exist, but one could use the Highly Adaptive Lasso (HAL) \citep{vanderlaan17}, which is an MLE over large class of functions but constraining the sectional variation norm so that it cannot overfit. It has been shown that the IPTW estimator using HAL for $\hat{g}$ is asymptotically efficient under some undersmoothing on the $L_1$-norm in the Highly Adaptive Lasso \citep{ertefaieetal23}. In other words, the undersmoothed HAL-MLE of $g_0$ is now always well-defined while preserving the asymptotic behavior the IPTW estimator would have under a nonparametric MLE (in case it would exist).

\subsection*{Defining consistency of the nonparametric and model-based bootstrap}
Suppose we are interested in evaluating the  performance of the IPTW estimator in a simulation study imitating the real study, or that we want to use a bootstrap to estimate the sampling distribution of the IPTW estimator in the actual data analysis. We could consider the following two types of bootstrap methods that differ in the choice of estimator of the true data distribution $P_0$ one samples from, which will be denoted with $\tilde{P}_{0,n}$ and ${\bf P}_{0,n}$, respectively.  Both methods fall in the category of model based bootstrap, contrary to the nonparametric bootstrap, since they do not correspond with sampling from the empirical probability  measure $P_n$. However, both use the empirical  measure for the covariate distribution, while using an estimator $P_{Y,n}$  of, or the actual, conditional distribution $P_{Y,0}$ of $Y$, given $W,A$.

Before we proceed we like to distinguish the two notions of consistency of the nonparametric bootstrap and consistency of a model based bootstrap for a given estimator $\hat{\Psi}(P_n)$ whose standardized estimator $n^{1/2}(\hat{\Psi}(P_n)-\Psi(P_0))$ is known to converge in distribution to a normal distribution $N(0,\sigma^2_0)$.
 \nl
{\bf Consistency of the nonparametric bootstrap:} Let $P_n^{\#}$ be the empirical measure of an i.i.d. sample from $P_n$. We say that the nonparametric bootstrap is consistent if, conditional on $(P_n:n\geq 1)$,  $n^{1/2}(\hat{\Psi}(P_n^{\#})-\hat{\Psi}(P_n))\Rightarrow_d N(0,\sigma^2_0)$. Note that the bootstrapped estimator is centered at the realization of the estimator on the original data defined by $P_n$, while the true parameter under the data distribution $P_n$, $\Psi(P_n)$, might not even be defined.

We note that for the above IPTW estimator for parametric MLE $\hat{g}(P_n)$, due to it being a compactly differentiable functional of the empirical measure $P_n$, the nonparametric bootstrap is consistent \citep{Gill89, vanderVaart&Wellner96}. We have also shown that the nonparametric bootstrap is also consistent for the IPTW estimator that uses an HAL-MLE $\hat{g}(P_n)$ \citep{cai&vanderlaan20}.

{\bf Consistency of a model based bootstrap:} Let ${\bf P}_n\in {\cal M}$ be a consistent estimator of $P_0$ that is an actual element of the statistical model ${\cal M}$. Let $P_n^{\#}$ be the empirical measure of an i.i.d. sample from ${\bf P}_n$. We say that the model based bootstrap is consistent if, conditional on $(P_n:n\geq 1)$, $n^{1/2}(\hat{\Psi}(P_n^{\#})-\Psi({\bf P}_n))\Rightarrow_d N(0,\sigma^2_0)$. Note that in this definition the target under sampling from ${\bf P}_n$ is $\Psi({\bf P}_n)$, the true target parameter under ${\bf P}_n$, and it is not $\hat{\Psi}(P_n)$ as in the nonparametric bootstrap. 
For a given estimator, one might have consistency of the nonparametric bootstrap and a model based bootstrap. 
While the nonparametric bootstrap is consistent for the IPTW estimator  we will show that  a particular type of  ${\bf P}_n$ (model based bootstrap II below) the model based bootstrap is inconsistent. 

For the sake of obtaining bootstrap based confidence intervals, the consistency of $n^{1/2}(\hat{\Psi}(P_n^{\#})-\hat{\Psi}(P_n))\Rightarrow_d N(0,\sigma^2_0)$ suffices. However, if the goal is to evaluate the performance of the estimator under a data distribution similar to $P_0$, including evaluating bias with respect to the truth, then one would use a model based bootstrap so that there is a well-defined target. 

\subsection*{Consistency of model based bootstrap that generates treatment}

{\bf Model based bootstrap I (generating treatment):} Let $\tilde{P}_{0,n}$ have density  given by $p_{0,n}(o)=p_n(w)g_n(a\mid w)p_{0,Y}(y\mid w,a)$, where $g_n=\hat{g}(P_n)$ is the MLE for model ${\cal G}$ as in the IPTW estimator.
To sample a $\tilde{O}=(\tilde{W},\tilde{A},\tilde{Y})$ from this data distribution $\tilde{P}_{0,n}$ one samples covariate $\tilde{W}$ from the empirical distribution of $W_1,\ldots,W_n$; one then simulates an $\tilde{A}$ from $g_n(\cdot\mid W)$; and then draws $\tilde{Y}$ from the conditional distribution $P_{Y,0}$ of $Y$, given $A,W$, under $P_0$.  Note $\Psi(\tilde{P}_{0,n})=P_n \bar{Q}_0$.  Let $\tilde{P}_n^{\#}$ be the empirical measure  of $\tilde{O}_i\sim_{iid} \tilde{P}_{0,n}$. one evaluates the performance of the estimator $\hat{\Psi}(P_n)$ by the sampling distribution of $n^{1/2}(\hat{\Psi}(\tilde{P}_n^{\#})-\Psi(\tilde{P}_{0,n}))$.

Let $\tilde{P}_n$ be the same data distribution as $\tilde{P}_{0,n}$ but with  $P_{0,Y}$ be replaced by  a uniformly consistent estimator $P_{Y,n}$ of $P_{Y,0}$, and let (again) $\tilde{P}_n^{\#}$ be the empirical measure of an i.i.d. sample from $\tilde{P}_n$. Under the data distribution $\tilde{P}_n$, the true target parameter is $\Psi(\tilde{P}_n)=P_n \bar{Q}_n$, where $\bar{Q}_n=E_{\tilde{P}_{n,Y}}(Y\mid W,A=1)$.

{\bf Consistency of model based bootstrap I:}
Conditional on $(P_n:n\geq 1)$, $n^{1/2}(\hat{\Psi}(\tilde{P_n}^{\#})-\Psi(\tilde{P}_{0,n}))$ converges in distribution to the same normal limit distribution $N(0,\sigma^2_0)$ as $n^{1/2}(\hat{\Psi}(P_n)-\Psi(P_0))$. The same statement holds under sampling from $\tilde{P}_n$: conditional on $(P_n:n\geq 1)$,  $n^{1/2}(\hat{\Psi}(\tilde{P}_n^{\#})-\Psi(\tilde{P}_n))\Rightarrow_d N(0,\sigma^2_0)$. 

\subsection*{Model based bootstrap that samples treatment}

{\bf Model based bootstrap II (sampling treatment):} We define the data distribution ${\bf P}_{0,n}$ defined by $(W,A)\sim P_n$ and conditional distribution of $Y$, given $A,W$, equals the conditional distribution $P_{Y,0}$ under $P_0$. So we could write
${\bf p}_{0,n}(o)=p_n(w,a)p_{0,Y}(y\mid w,a)$  with $p_n$ the discrete probability distribution with mass $1/n$ on each $(W_i,A_i)$, $i=1,\ldots,n$. One would define the true target as  $\Psi({\bf P}_{0,n})=P_n \bar{Q}_0$, which represents the true parameter value under ${\bf P}_{0,n}$ (however, as noted earlier, strictly speaking $\Psi({\bf P}_{0,n})$ is  ill-defined). Similarly, let ${\bf P}_n$ be the same data distribution as ${\bf P}_{0,n}$ but with $P_{0,Y}$ replaced by a uniformly consistent estimator $P_{Y,n}$. So ${\bf p}_n(o)=p_n(w,a)p_{Y,n}(y\mid w,a)$. In this case one would define the true target as $\Psi({\bf P}_n)=P_n \bar{Q}_n$.
 In both cases, let $P_n^{\#}$ be the empirical measure of $n$ i.i.d. copies $O_i^{\#}\sim_{iid} {\bf P}_{0,n}$ or $O_i^{\#}\sim_{iid}{\bf P}_n$,  $i=1,\ldots,n$. 
 In a bootstrap one evaluates the performance of the estimator by the sampling distribution of $n^{1/2}(\hat{\Psi}(P_n^{\#})-\Psi({\bf P}_n))$.
 
 \subsection*{Inconsistency of the model based bootstrap sampling treatment}
 {\bf Key problem with the ${\bf P}_{0,n}$-bootstrap:} Even though $\Psi({\bf P}_{0,n})=P_n \bar{Q}_0$ is a well-defined target, note that
 $E_{P_n}E_{P_{Y,0}}(Y\mid A=1,W)$ is ill defined due to $P_n(A=1,W=W_i)$ equals zero for any observation $i$ with $A_i=0$ so that the conditional expectation is not defined. This already is an indication of a potential  problem with this choice of data distribution ${\bf P}_{0,n}$.
 A key difference between the two choices $\tilde{P}_{0,n}$ and ${\bf P}_{0,n}$ is that
 \[
 \hat{\Psi}(\tilde{P}_{0,n})=\tilde{P}_{0,n}AY/\hat{g}(\tilde{P}_{0,n})=P_n g_n/g_n \bar{Q}_0=P_n \bar{Q}_0=\Psi(\tilde{P}_{0,n}),\]
 while
  \[
 \hat{\Psi}({\bf P}_{0,n})={\bf P}_{0,n} Y A/\hat{g}({\bf P}_{0,n})=
 P_n \bar{Q}_0 A/\hat{g}(P_n)\not = P_n \bar{Q}_0.\]
 This inconsistent definition of the estimator $\hat{\Psi}({\bf P}_{0,n})\not =\Psi({\bf P}_{0,n})$ causes the bias  in the ${\bf P}_{0,n}$-based bootstrap, as follows from the proof below.

To establish consistency of this bootstrap method II one would  need to show that, conditional on $(P_n:n\geq 1)$,  $n^{1/2}(\hat{\Psi}(P_n^{\#})-\Psi({\bf P}_{0,n}))$ converges to the same normal limit distribution $N(0,\sigma^2_0)$ of $n^{1/2}(\hat{\Psi}(P_n)-\Psi(P_0))$. The same proof below applies to the bootstrap distribution ${\bf P}_n$ by just replacing $P_{Y,0}$ by $P_{Y,n}$. We note that we still have $n^{1/2}(\hat{\Psi}(P_n^{\#})-\hat{\Psi}({\bf P}_{0,n}))\Rightarrow_d N(0,\sigma^2_0)$. The fact that the nonparametric bootstrap works also implies that this  latter model based bootstrap works. 

Conditionally on $(P_n:n\geq 1)$,
we have
\[
\hat{\Psi}(P_n^{\#})-\Psi({\bf P}_{0,n})=\{\hat{\Psi}(P_n^{\#})-\hat{\Psi}({\bf P}_{0,n})\}+
\hat{\Psi}({\bf P}_{0,n})-\Psi({\bf P}_{0,n}),\]
where the first term  converges to $N(0,\sigma^2_0)$ by consistency of this model based bootstrap for $\hat{\Psi}$. The second term is fixed given $P_n$ and thus represents the bias term $B_n$ which can be worked out as follows:
\begin{eqnarray*}
B_{n}&=&\hat{\Psi}({\bf P}_{0,n})-\Psi({\bf P}_{0,n})\\
&=& {\bf P}_{0,n} AY/\hat{g}(P_n)-P_n \bar{Q}_0\\
&=& P_n A \bar{Q}_0/g_n -P_n \bar{Q}_0\\
&=& P_n \bar{Q}_0/g_n(A-g_n)\\
&=&(P_n-P_0)\bar{Q}_0/g_n(A-g_n)+P_0 \bar{Q}_0/g_n(A-g_n)\\
&=&(P_n-P_0)\bar{Q}_0/g_0(A-g_0)+o_P(n^{-1/2})-P_0 \bar{Q}_0/g_0(g_n-g_0)+o_P(n^{-1/2})\\
&=&(P_n-P_0)\{ \bar{Q}_0/g_0(A-g_0)-D_{{\cal G},P_0}\}+o_P(n^{-1/2}),
\end{eqnarray*}
where we used that $P_0\bar{Q}_0/g_0(g_n-g_0)=(P_n-P_0)D_{{\cal G},P_0}+o_P(n^{-1/2})$.
Let $D_{CAR,P_0}\equiv \bar{Q}_0/g_0(A-g_0)$ and  $D_{1,P_0}\equiv D_{CAR,P_0}-D_{{\cal G},P_0}$.

Note that marginally $n^{1/2}B_{n}$ converges to a normal distribution with mean zero and variance $\sigma^2_1\equiv P_0\{D_{1,P_0}\}^2$, but conditional on $P_n$, this means that $n^{1/2}B_{n}$ represents a bias of order $O(1)$. 

\subsection*{Theorem stating inconsistency of model based bootstrap sampling treatment}
\begin{theorem}
Consider a sample of $n$ i.i.d. copies of $(W,A,Y)\sim P_0$ and let $P_n$ be the empirical measure of $O_1,\ldots,O_n$. Let $g_0(1\mid W)=P_0(A=1\mid W)$ and $\bar{Q}_0(W)=E_0(Y\mid A=1,W)$. 
Suppose the target parameter of interest $\Psi(P_0)=E_0\bar{Q}_0(W)$. Assume $g_0(1\mid W)>\delta>0$, $P_0$-a.e.
Let $\hat{g}(P_n)$ be a maximum likelihood estimator of $g_0$ according to a correctly specified parametric model ${\cal G}$ with tangent space $T_{\cal G}(P_0)$ at $P_0$. The statistical model for $P_0$ is given by ${\cal M}({\cal G})$ that only assumes $g_0\in {\cal G}$.
A possible estimator of $\psi_0$ is given by the IPTW estimator $\psi_n=\hat{\Psi}(P_n)=E_{P_n} AY/\hat{g}(P_n)$. 

We have that $\hat{\Psi}(P_n)$ is an asymptotically linear estimator of $\Psi(P_0)$ with influence curve $D_{IPTW,P_0}\equiv D_{P_0}-D_{{\cal G},P_0}$, where $D_{P_0}(O)=AY/g_0(W)-\psi_0$ and $D_{{\cal G},P_0}=\Pi(D_{P_0}\mid T_{{\cal G}}(P_0))$ denotes the projection of $D_{P_0}$ onto sub-Hilbert space $T_{{\cal G}}(P_0)\subset L^2_0(P_0)$ of $L^2_0(P_0)$ with inner product $\langle h_1,h_2\rangle_{P_0}P_0 h_1h_2$ the covariance operator under $P_0$. Let $\sigma^2_0\equiv P_0\{D_{IPTW,P_0}\}^2$.
Thus, $n^{1/2}(\hat{\Psi}(P_n)-\Psi(P_0))\Rightarrow_d N(0,\sigma^2_0)$. 

{\bf Model based bootstrap II, sampling treatment:} We define the data distribution ${\bf P}_{0,n}$ defined by $(W,A)\sim P_n$ and conditional distribution of $Y$, given $A,W$, equals the conditional distribution $P_{Y,0}$ under $P_0$. So the density of $P_{0,n}$ is given by ${\bf p}_{0,n}(o)=p_n(w,a)p_{0,Y}(y\mid w,a)$  with $p_n$ the discrete probability distribution with mass $1/n$ on each $(W_i,A_i)$, $i=1,\ldots,n$. Suppose we define the true target as $\Psi({\bf P}_{0,n})=P_n \bar{Q}_0$ as the true parameter value under $P_{0,n}$ (even though it is not well defined). Let $P_n^{\#}$ be the empirical measure of $n$ i.i.d. copies $O_i^{\#}\sim_{iid} P_{0,n}$, $i=1,\ldots,n$. 

Conditional on $(P_n: n\geq 1)$, we have
\[
\hat{\Psi}(P_n^{\#})-\Psi({\bf P}_{0,n})=\{\hat{\Psi}(P_n^{\#})-\hat{\Psi}({\bf P}_{0,n})\}+B_{n},\]
where the bias term is given by
$B_{n}=\hat{\Psi}({\bf P}_{0,n})-\Psi({\bf P}_{0,n})$,
while the first term scaled by $n^{1/2}$ converges to $N(0,\sigma^2_0)$ (the same limit distribution as the standardized IPTW estimator under sampling from $P_0$). 
Let $D_{1,P_0}\equiv D_{CAR,P_0}-D_{{\cal G},P_0}$. We have, conditionally on $(P_n:n\geq 1)$, \[
B_{n}=(P_n-P_0)D_{1,P_0}+o(n^{-1/2}).\]

As a consequence, the sampling distribution of $n^{1/2}(\hat{\Psi}(P_n^{\#})-\Psi({\bf P}_{0,n}))$  approximates $N(n^{1/2}B_{n},\sigma^2_0)=N(0,\sigma^2_0)+n^{1/2}B_{n}$, where $n^{1/2}B_{n}$ represents the realization of a $N(0,\sigma^2_1)$ random variable.
\end{theorem}
So we can conclude that  bootstrap method II is inconsistent for estimating the sampling distribution of the IPTW estimator. Thus, a  plasma simulation based on method II would demonstrate that the standardized IPTW estimator $n^{1/2}(\hat{\Psi}(P_n^{\#})-\Psi({\bf P}_{0,n}) )$ has a non-disappearing bias $n^{1/2}B_{n}\sim N(0,\sigma^2_1)$ in its normal limit distribution, while under sampling from an appropriate data distribution such as $P_0$ its sampling distribution would be well approximated by the mean zero normal distribution $N(0,\sigma^2_0)$. 
The exception occurs if $\hat{g}(P_n)$ is a nonparametric MLE only assuming the CAR assumption, as can be approximated with the HAL-MLE, in which case $D_{{\cal G},P}=D_{CAR,P}$ so that $D_{1,P}=0$.

\subsection*{Inconsistency of model based bootstrap II to propensity score adjusted  G-comp estimators}
Define $Q_{g_n,0}^r=E_0(Y\mid A=1,g_n(W))$ and let 
$Q_n^r=\hat{E}(Y\mid A=1,\hat{g}(P_n))$ be an HAL-MLE regressing $Y$ on $A=1$ and $\hat{g}(P_n)(W)$.
We also define $Q_0^r=Q_{g_0,0}^r=E_0(Y\mid A=1,g_0(W))$. 
Let $\hat{\Psi}(P_n)=P_n Q_n^r$ be the estimator of $\Psi(P_0)=P_0 Q_0=P_0 Q_0^r$.
We have
\[
\begin{array}{l}
P_n \hat{E}(Y\mid A=1,\hat{g}(P_n))-P_0Q_0^r=(P_n-P_0)(Q_0^r-\Psi(Q_0))+
P_0 (Q_{g_n,n}^r-Q_{g_0,0}^r)\\
=(P_n-P_0)(Q_0^r-\psi_0)+P_0(Q_{g_n,n}^r-Q_{g_n,0}^r)+P_0(Q_{g_n,0}^r-Q_{g_0,0}^r)\\
=(P_n-P_0)(Q_0^r-\psi_0)+(P_n-P_0)\{A/P_0(A=1\mid g_n(W))(Y-Q_0^r(W))\\
\hfill -(P_n-P_0)D_{{\cal G},P_0}+o_P(n^{-1/2})\\
=(P_n-P_0)\{Q_0^r-\psi_0+A/g_0(Y-Q_0^r(W))-D_{{\cal G},P_0})+o_P(n^{-1/2}),
\end{array}
\]
where $D_{{\cal G},P_0}$ is the projection of $A/g_0(Y-Q_0^r)$ onto $T_{\cal G}(P_0)$.
Here we relied on showing that $P_0(Q_{g_n,0}^r-Q_{g_0,0}^r)$ is asymptotically linear, which will require smoothness assumptions, which is beyond the scope of this Appendix. 
Let  $D^r_{P_0}(O)=A/g_0(W)(Y-Q_0^r(W))+Q_0^r(W)-\Psi(P_0)$.
So we have that $\hat{\Psi}(P_n)$ is asymptotically linear with influence curve $D^r_{P_0}-D_{{\cal G},P_0}$.

Note $\hat{\Psi}({\bf P}_{0,n})= P_n E_0(Y\mid A=1,g_n(W))$.
We have asymptotic consistency of the bootstrap for $n^{1/2}(\hat{\Psi}(P_n^{\#})-\hat{\Psi}({\bf P}_{0,n}))$ to the same limiting distribution as $n^{1/2}(\hat{\Psi}(P_n)-\hat{\Psi}(P_0))$ provided above.
We conclude that the bias term is given by $B_{1,n}=\hat{\Psi}({\bf P}_{0,n})-P_n \bar{Q}_0$.
Specifically, 
\begin{eqnarray*}
B_{1,n}&=&P_n(Q^r_{g_n,0}-\bar{Q}_0)\\
&=&(P_n-P_0)(Q^r_{g_n,0}-Q_0)+P_0(Q_{g_n,0}^r-Q_0)\\
&=&
(P_n-P_0)\{Q_{g_0,P_0}^r-\bar{Q}_0\}+P_0\{Q_{g_n,0}^r-Q_{g_0,0}^r\}+o_P(n^{-1/2})\\
&=&(P_n-P_0)\{Q_{g_0,P_0}^r-\bar{Q}_0\}-(P_n-P_0)D_{{\cal G},P_0}+o_P(n^{-1/2}).
\end{eqnarray*}

Let's now consider the case that we sample from $\tilde{P}_{0,n}$ (generate treatment). We have $\hat{\Psi}(\tilde{P}_{0,n})=P_n E_0(Y\mid A=1,g_n(W))=P_n \bar{Q}_0$. This is due to the general result that $E_P E_P(Y\mid g_P(W),A=1)=E_PE_P(Y\mid A=1,W)$ applied to $P=\tilde{P}_{0,n}=p_{W,n}g_nq_{Y,0}$.
So now $B_{1,n}=\hat{\Psi}({\bf P}_{0,n})-P_n\bar{Q}_0=0$. So this bias is not present in the model based bootstrap I method.

Therefore, we can conclude that also for propensity score adjustment estimators, assuming the regularity conditions that makes these estimators asymptotically linear and smooth enough for the bootstrap to work for $\hat{\Psi}(P_n)$, the model based bootstrap based on ${\bf P}_{0,n}$ is inconsistent with a specified bias term $B_{1,n}$, while the model based bootstrap based on sampling from $\tilde{P}_{0,n}$ is consistent.

We should expect the same issue for propensity score matching estimators. These propensity score adjustment estimators might suffer more due the asymptotic consistency of the nonparametric bootstrap already relying on some smoothness conditions (just as machine learning or nonparametric regression based estimators having issues with a nonparametric bootstrap, and that the ${\bf P}_{0,n}$-bootstrap might still have issues due to the distribution of $(W,A)$ being discrete. 

Similarly, double robust estimators such as TMLE or A-IPCW will also show a bias in the ${\bf P}_{0,n}$ bootstrap if $\bar{Q}_n$ is inconsistent.

\section{Additional details for the synthetic data plasmode simulations} \label{sec:supp_synthetic}
\subsection{Details for the simulation set-up }

For Scenario 4b, we estimated the true values for the MSM by the following steps:
\begin{enumerate}
\item [Step 1.] Generate data $O=(\textbf{W}$, A, Y) according to the DGM for $n = 100,000$
\item [Step 2.] Evaluate $Q(0,\tilde{w}) = P(Y=1|A = 0, \widetilde{W}=\tilde{w})$ and $Q(1,\tilde{w}) = P(Y=1|A = 1, \widetilde{W}=\tilde{w})$ (known from DGM)
\item [Step 3.] Create a stacked dataset with $2n$ observations, with the first $n$ observations having outcome $Y = Q(0,\widetilde{W})$, treatment $A = 0$, and covariates $W$, and the second set of $n$ observations having outcome $Y = Q(1,\widetilde{W})$, treatment $A = 1$, and covariates $\widetilde{W}$.
\item [Step 4.] Regress the stacked dataset onto the MSM and record the estimated coefficients.
\end{enumerate}

 To establish the truth for Scenario 4b, Steps 1-4 were repeated 500 times and the true MSM parameter values were estimated as the mean values for each coefficient, $\mathbf{\gamma} =  (-2.697,  1.084,  0.005,  0.079 , 0.492)$, where 1.084 is the logcOR. 

\subsection{Supplemental Tables and Figures for the synthetic data plasmode simulations}

 \begin{figure}[htb] 
   \centering
     \caption{Nominal coverage of 95\% confidence intervals for (a) Simulation Scenario 1: continuous outcome and (b) Simulation Scenario 2 binary outcome } 
     
     \subfloat[Scenario 1: Continuous outcome]{\includegraphics[width=5.75in]{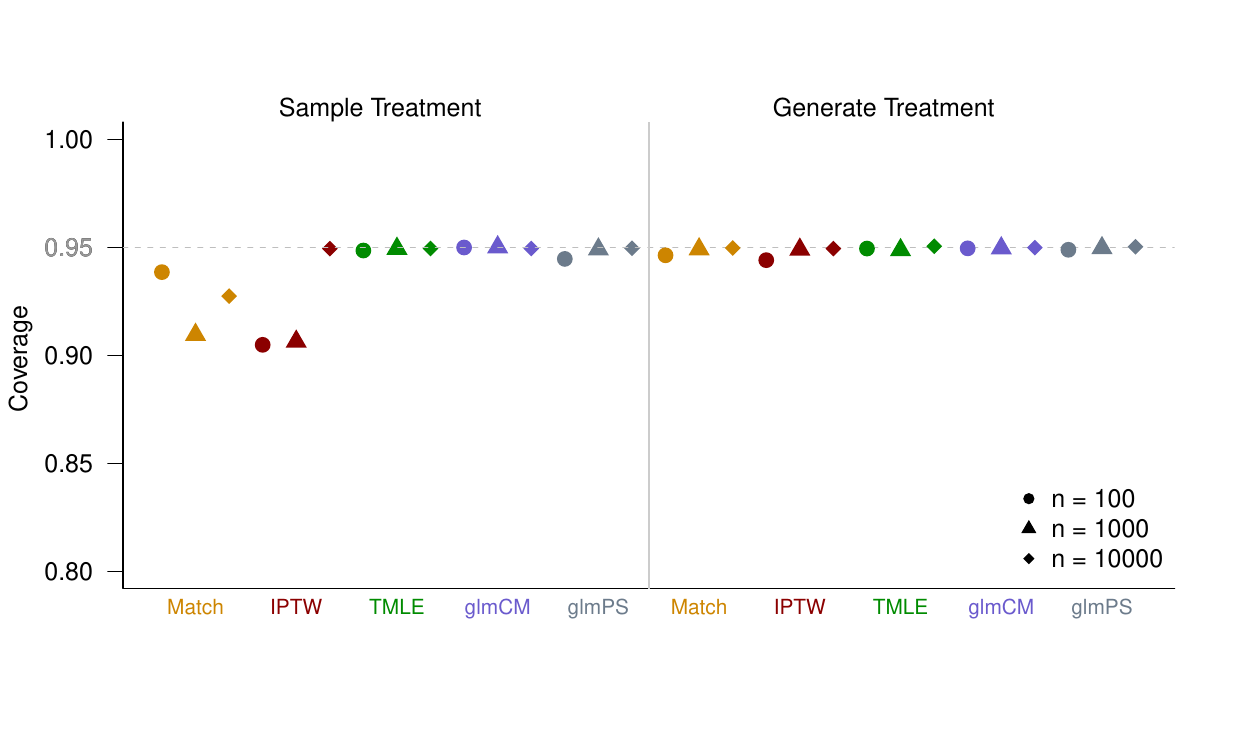}  \label{fig:coverageCont}}
          
    \subfloat[Scenario 2: Binary outcome]{
    \includegraphics[width=5.75in]{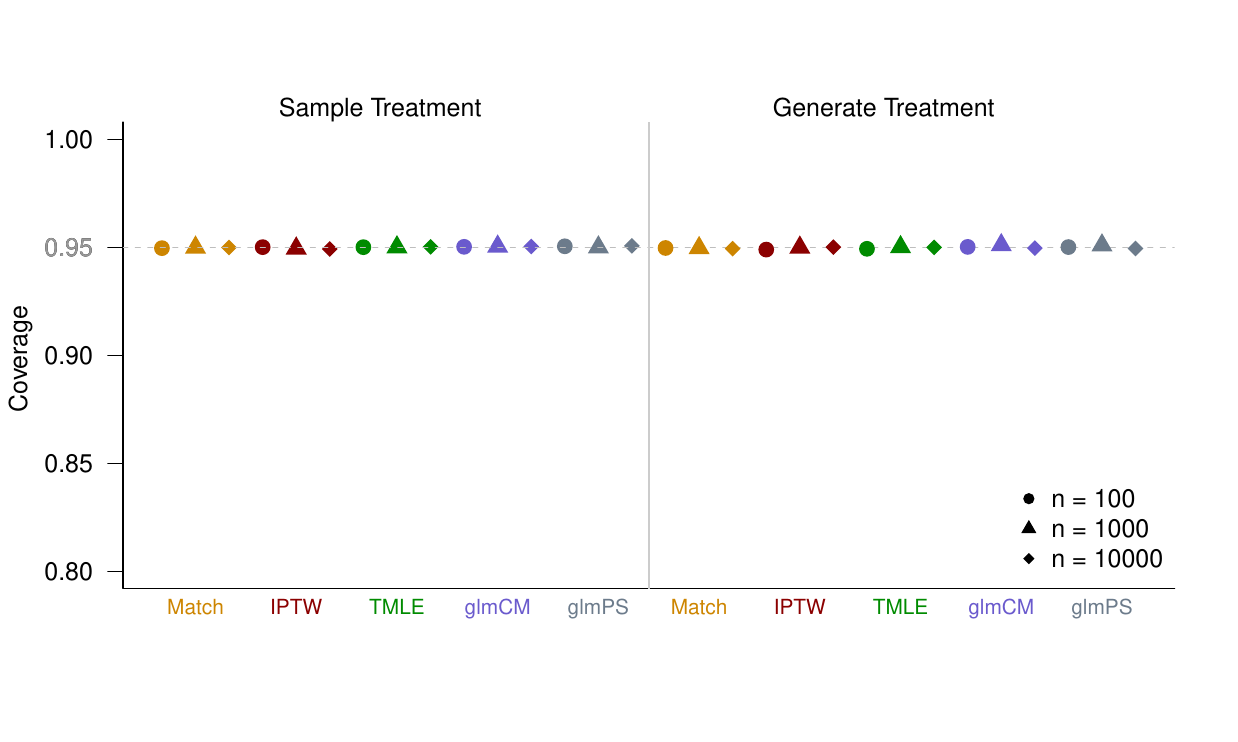}  \label{fig:coverageBin}
}
\end{figure}

\clearpage

\begin{table}[!htbp]
\caption{Simulation Scenario 1a: RCT, Estimate ATE for continuous outcome, $n=1000$\label{tab:sim1an1000}} 
\begin{center}
\begin{tabular}{lrrrrcrrrr}
\hline\hline
\multicolumn{1}{l}{\bfseries }&\multicolumn{4}{c}{\bfseries Sample Treatment}&\multicolumn{1}{c}{\bfseries }&\multicolumn{4}{c}{\bfseries Generate Treatment}\tabularnewline
\cline{2-5} \cline{7-10}
\multicolumn{1}{l}{}&\multicolumn{1}{c}{\% Bias}&\multicolumn{1}{c}{SE}&\multicolumn{1}{c}{RMSE}&\multicolumn{1}{c}{Bias:SE}&\multicolumn{1}{c}{}&\multicolumn{1}{c}{\% Bias}&\multicolumn{1}{c}{SE}&\multicolumn{1}{c}{RMSE}&\multicolumn{1}{c}{Bias:SE}\tabularnewline
\hline
Unadj&$10.737$&$0.470$&$0.719$&$0.457$&&$-0.107$&$0.471$&$0.686$&$0.005$\tabularnewline
Match&$ 0.001$&$0.256$&$0.506$&$0.000$&&$ 0.048$&$0.193$&$0.439$&$0.005$\tabularnewline
IPTW&$10.737$&$0.470$&$0.719$&$0.457$&&$-0.107$&$0.471$&$0.686$&$0.005$\tabularnewline
TMLE&$-0.235$&$0.063$&$0.252$&$0.074$&&$-0.078$&$0.063$&$0.252$&$0.024$\tabularnewline
glmCM&$-0.002$&$0.063$&$0.251$&$0.001$&&$-0.013$&$0.063$&$0.252$&$0.004$\tabularnewline
glmPS&$10.737$&$0.470$&$0.719$&$0.457$&&$-0.107$&$0.471$&$0.686$&$0.005$\tabularnewline
\hline
\multicolumn{10}{l}{\footnotesize RCT: randomized controlled trial; ATE: average treatment effect; SE: Standard Error; RMSE: root mean} \tabularnewline
\multicolumn{10}{l}{\footnotesize  squared error; Unadj: unadjusted; IPTW: inverse probability of treatment weighted; TMLE: targeted}\tabularnewline
\multicolumn{10}{l} {\footnotesize  maximum likelihood estimation; glmCM: generalized linear model correct model; glmPS: generalized}\tabularnewline
\multicolumn{10}{l}{\footnotesize linear model propensity score} \tabularnewline
\end{tabular}\end{center}
\end{table}

 \begin{table}[!htbp]
\caption{Simulation Scenario 2a: RCT,Estimate ATE for binary outcome, $n=1000$\label{tab:sim2an1000}} 
\begin{center}
\begin{tabular}{lrrrrcrrrr}
\hline\hline
\multicolumn{1}{l}{\bfseries }&\multicolumn{4}{c}{\bfseries Sample Treatment}&\multicolumn{1}{c}{\bfseries }&\multicolumn{4}{c}{\bfseries Generate Treatment}\tabularnewline
\cline{2-5} \cline{7-10}
\multicolumn{1}{l}{}&\multicolumn{1}{c}{\% Bias}&\multicolumn{1}{c}{SE}&\multicolumn{1}{c}{RMSE}&\multicolumn{1}{c}{Bias:SE}&\multicolumn{1}{c}{}&\multicolumn{1}{c}{\% Bias}&\multicolumn{1}{c}{SE}&\multicolumn{1}{c}{RMSE}&\multicolumn{1}{c}{Bias:SE}\tabularnewline
\hline
Unadj&$ 3.652$&$0.028$&$0.171$&$0.281$&&$-0.022$&$0.028$&$0.169$&$0.002$\tabularnewline
Match&$-0.011$&$0.032$&$0.179$&$0.001$&&$-0.028$&$0.030$&$0.174$&$0.002$\tabularnewline
IPTW&$ 3.652$&$0.028$&$0.171$&$0.281$&&$-0.022$&$0.028$&$0.169$&$0.002$\tabularnewline
TMLE&$ 0.020$&$0.028$&$0.167$&$0.002$&&$-0.024$&$0.028$&$0.167$&$0.002$\tabularnewline
glmCM&$ 0.019$&$0.028$&$0.166$&$0.001$&&$-0.025$&$0.028$&$0.167$&$0.002$\tabularnewline
glmPS&$ 3.652$&$0.028$&$0.171$&$0.281$&&$-0.022$&$0.028$&$0.169$&$0.002$\tabularnewline
\hline
\multicolumn{10}{l}{\footnotesize RCT: randomized controlled trial; ATE: average treatment effect; SE: Standard Error; RMSE: root mean} \tabularnewline
\multicolumn{10}{l}{\footnotesize  squared error; Unadj: unadjusted; IPTW: inverse probability of treatment weighted; TMLE: targeted}\tabularnewline
\multicolumn{10}{l} {\footnotesize  maximum likelihood estimation;glmCM: generalized linear model correct model; glmPS: generalized}\tabularnewline
\multicolumn{10}{l}{\footnotesize linear model propensity score} \tabularnewline
\end{tabular}\end{center}
\end{table}

 \begin{figure}[htbp] 
   \centering
      \caption{Estimator bias and coverage under each sampling scheme for simulation Scenario 1a: Continuous outcome in a randomized controlled trial setting (n=1000) }
\subfloat[Estimator bias under each plasmode sampling scheme]{
\includegraphics[width=5.75in]{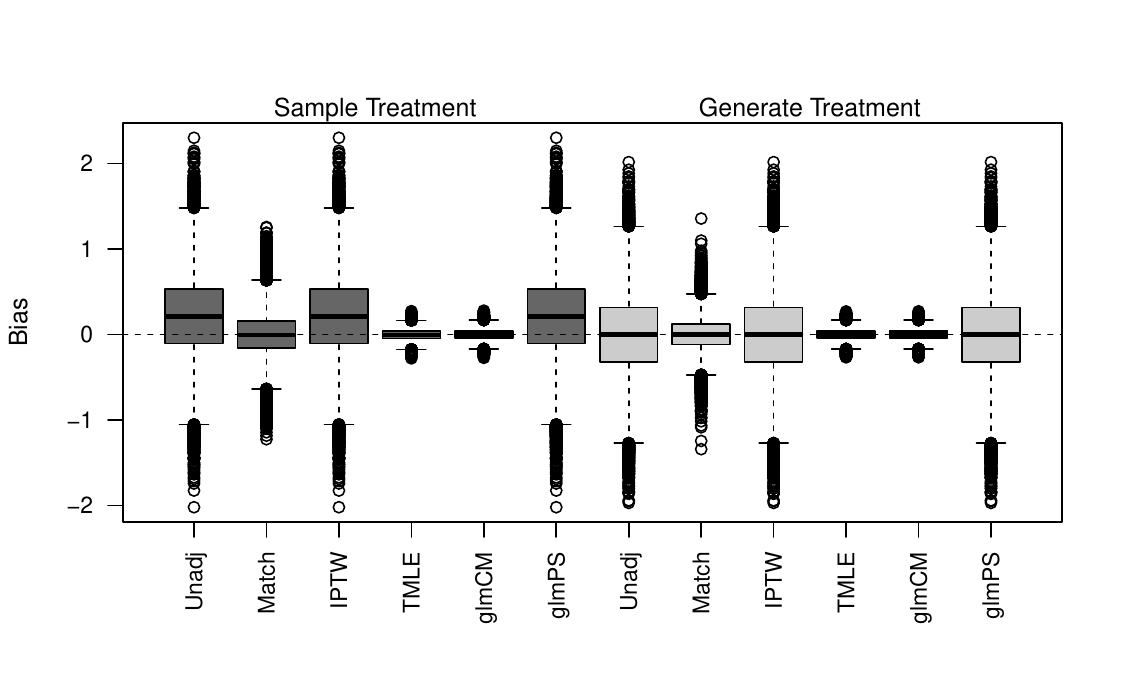} 
   \label{fig:sim1abox} }
   
      \subfloat[Coverage under each plasmode sampling scheme]{
\includegraphics[width=5.75in]{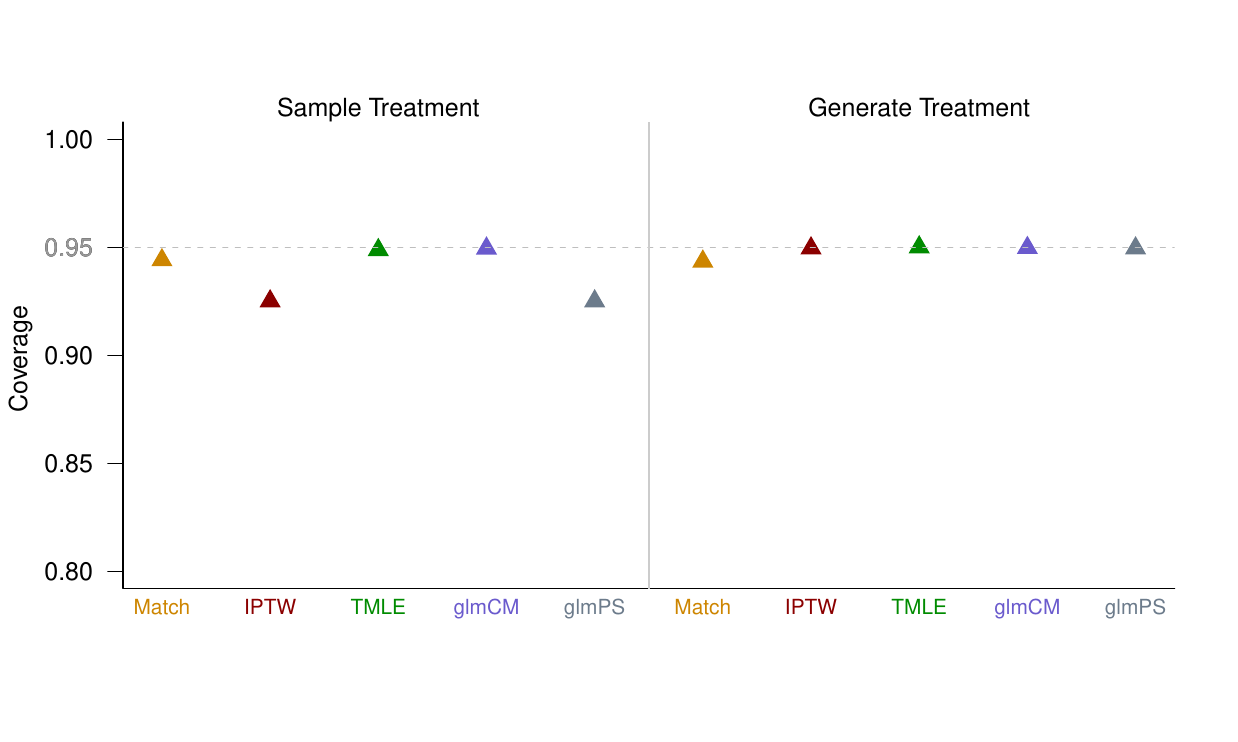} 
   \label{fig:sim1aCov}}
\end{figure}

 \begin{figure}[htpb] 
   \centering
      \caption{Estimator bias and coverage under each sampling scheme for simulation Scenario 2a: Binary outcome in a randomized controlled trial setting (n=1000)} 
      
\subfloat[Estimator bias under each plasmode sampling scheme]{\includegraphics[width=5.75in]{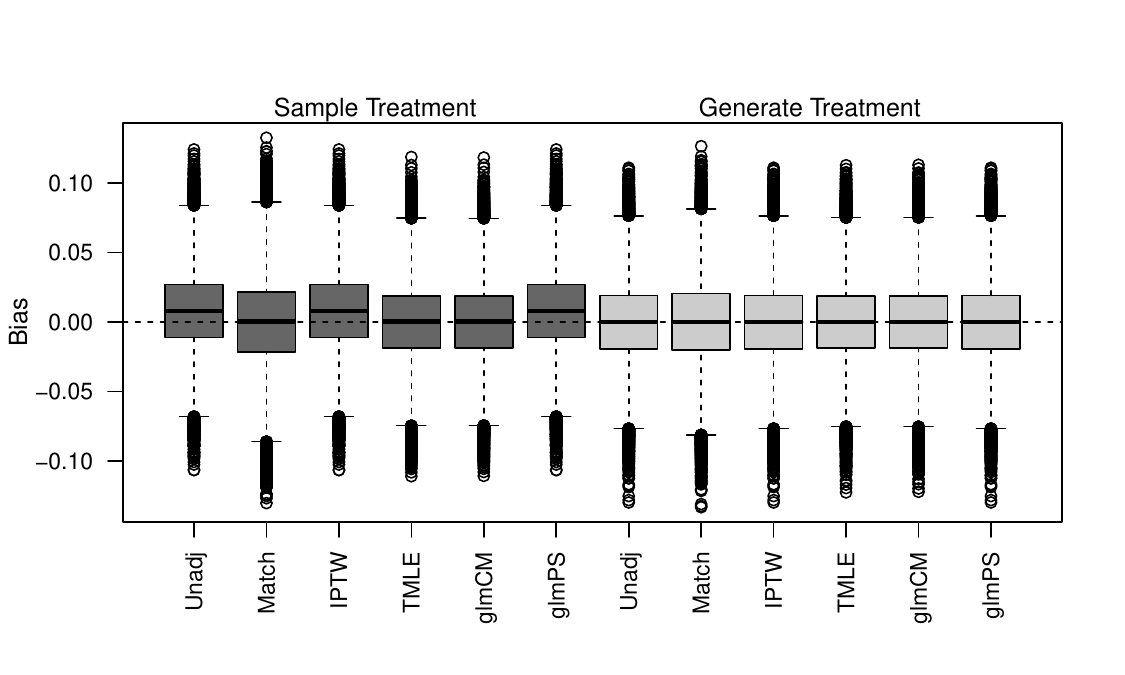} 
   \label{fig:sim2abox}}

      \subfloat[Coverage under each plasmode sampling scheme]{
\includegraphics[width=5.75in]{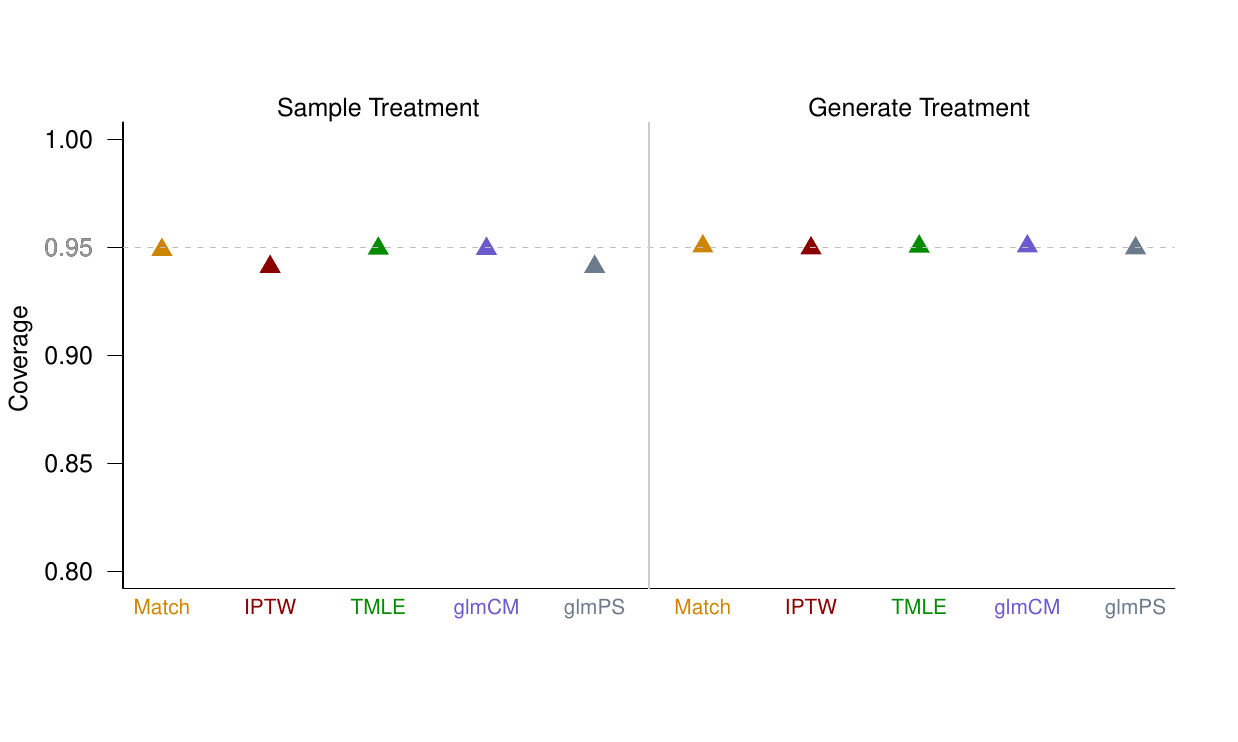} 
   \label{fig:sim2aCov} }
\end{figure}

\clearpage{}

\begin{table}[!htbp]
\caption{Simulation Scenario 3: Estimate RR for rare binary outcome, $n=10000$\label{tab:sim3RRn10000}} 
\begin{center}
\begin{tabular}{lrrrrcrrrr}
\hline\hline
\multicolumn{1}{l}{\bfseries }&\multicolumn{4}{c}{\bfseries Sample Treatment}&\multicolumn{1}{c}{\bfseries }&\multicolumn{4}{c}{\bfseries Generate Treatment}\tabularnewline
\cline{2-5} \cline{7-10}
\multicolumn{1}{l}{}&\multicolumn{1}{c}{\% Bias}&\multicolumn{1}{c}{SE}&\multicolumn{1}{c}{RMSE}&\multicolumn{1}{c}{Bias:SE}&\multicolumn{1}{c}{}&\multicolumn{1}{c}{\% Bias}&\multicolumn{1}{c}{SE}&\multicolumn{1}{c}{RMSE}&\multicolumn{1}{c}{Bias:SE}\tabularnewline
\hline
Unadj&$-40.788$&$0.036$&$0.145$&$3.967$&&$-32.628$&$0.039$&$0.119$&$2.919$\tabularnewline
Match&$ -9.993$&$0.072$&$0.080$&$0.479$&&$  0.556$&$0.071$&$0.071$&$0.027$\tabularnewline
IPTW&$-11.885$&$0.056$&$0.069$&$0.737$&&$  0.405$&$0.061$&$0.061$&$0.023$\tabularnewline
TMLE&$  0.113$&$0.049$&$0.049$&$0.008$&&$  0.225$&$0.049$&$0.049$&$0.016$\tabularnewline
glmCM&$  0.228$&$0.046$&$0.046$&$0.017$&&$  0.224$&$0.045$&$0.045$&$0.017$\tabularnewline
glmPS&$-14.844$&$0.052$&$0.073$&$0.992$&&$ -2.038$&$0.056$&$0.057$&$0.125$\tabularnewline
\hline
\multicolumn{10}{l}{\footnotesize RR: relative risk; SE: Standard Error; RMSE: root mean squared error; Unadj: unadjusted;} \tabularnewline
\multicolumn{10}{l}{\footnotesize IPTW: inverse probability of treatment weighted; TMLE: targeted maximum likelihood estimation;}\tabularnewline
\multicolumn{10}{l} {\footnotesize glmCM: generalized linear model correct model; glmPS: generalized linear model propensity score}\tabularnewline
\end{tabular}\end{center}
\end{table}

 \begin{figure}[htpb] 
   \centering
      \caption{Simulation Scenario 3 (rare binary outcome) Coverage for RR} 
\includegraphics[width=6in]{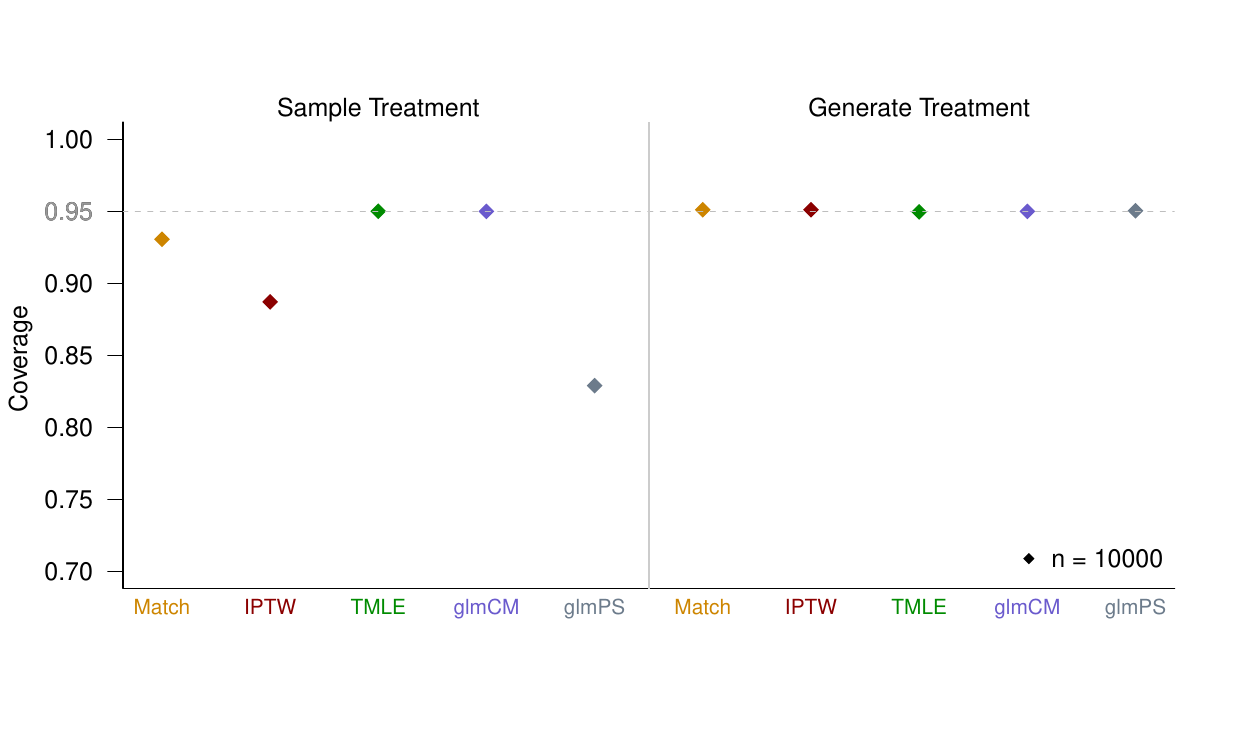} 
   \label{fig:sim3CovRR}
\end{figure}

\clearpage

\begin{table}[!htbp]
\caption{Simulation Scenario 4a: Estimate logcOR when MSM equivalent to true outcome regression} \label{tab:sim4a-alln} 
\begin{center}
\begin{tabular}{lrrrrcrrrr}
\hline\hline
\multicolumn{1}{l}{\bfseries }&\multicolumn{4}{c}{\bfseries Sample Treatment}&\multicolumn{1}{c}{\bfseries }&\multicolumn{4}{c}{\bfseries Generate Treatment}\tabularnewline
\cline{2-5} \cline{7-10}
\multicolumn{1}{l}{}&\multicolumn{1}{c}{\% Bias}&\multicolumn{1}{c}{SE}&\multicolumn{1}{c}{RMSE}&\multicolumn{1}{c}{Bias:SE}&\multicolumn{1}{c}{}&\multicolumn{1}{c}{\% Bias}&\multicolumn{1}{c}{SE}&\multicolumn{1}{c}{RMSE}&\multicolumn{1}{c}{Bias:SE}\tabularnewline
\hline
$n=100$&$17.574$&$1.055$&$1.072$&$0.181$&&$20.760$&$1.056$&$1.079$&$0.213$\tabularnewline
$n=1000$&$ 2.837$&$0.231$&$0.233$&$0.133$&&$ 2.778$&$0.232$&$0.234$&$0.130$\tabularnewline
$n=10000$&$ 1.643$&$0.072$&$0.074$&$0.248$&&$ 1.621$&$0.071$&$0.073$&$0.246$\tabularnewline
\hline
\multicolumn{10}{l}{\footnotesize logcOR: log conditional odds ratio; MSM: marginal structural model; SE: Standard Error; RMSE: root} \tabularnewline
\multicolumn{10}{l}{\footnotesize mean squared error; Unadj: unadjusted; IPTW: inverse probability of treatment weighted; TMLE: targeted}\tabularnewline
\multicolumn{10}{l} {\footnotesize maximum likelihood estimation; glmCM: generalized linear model correct model; glmPS: generalized linear}\tabularnewline
\multicolumn{10}{l}{\footnotesize model propensity score}
\end{tabular}\end{center}
\end{table}

\clearpage


\section{Additional set-up details for the electronic health records-based plasmode simulations}\label{sec:supp_kpwa_analysis}

We present regression model coefficients for creating plasmode datasets in Table~\ref{tab:original_coefficients}.

\begin{table}

\caption{Logistic regression model coefficients fit to the original KPWA dataset with 50,337 individuals. These coefficients differ from \citet{williamson2024} because Charlson score was originally treated as an ordinal variable, and here it is not. The second column gives coefficients in the propensity score model for treatment with psychotherapy (PT). The third column gives coefficients from the logistic regression model fit to the data for the 5-year self-harm or hospitalization (SH/HOSP) composite outcome. The fourth and fifth column give coefficients for generating plasmode data with a 15\% and 5\% outcome rate, respectively.\label{tab:original_coefficients}}
\centering
\begin{tabular}[t]{lrrrr}
\toprule
Variable & Receipt of PT & 5-year SH/HOSP & 15\% outcome & 5\% outcome\\
\midrule
Intercept & 2.361 & -2.063 & -1.320 & -2.350\\
Psychotherapy & NA & -0.206 & -1.000 & -3.100\\
Female sex & -0.238 & 0.360 & 0.360 & 0.360\\
Age at initiation & -0.030 & -0.060 & -0.060 & -0.060\\
Charlson 1 & -0.041 & 0.176 & 0.176 & 0.176\\
Charlson 2 & 0.084 & 0.953 & 0.953 & 0.953\\
Charlson 3+ & 0.907 & 1.988 & 1.988 & 1.988\\
Alcohol use disorder & 0.242 & 0.842 & 0.842 & 0.842\\
Anxiety disorder & 0.454 & 0.096 & 0.096 & 0.096\\
Prior self-harm & 0.145 & 1.960 & 1.960 & 1.960\\
Prior hospitalization with MH diagnosis & -0.320 & 0.914 & 0.914 & 0.914\\
PHQ8: 6--10 & -0.878 & -0.026 & -0.026 & -0.026\\
PHQ8: 11--15 & -1.674 & 0.209 & 0.209 & 0.209\\
PHQ8: 16--20 & -2.074 & 0.338 & 0.338 & 0.338\\
PHQ8: 21--24 & -2.126 & 0.349 & 0.349 & 0.349\\
PHQ9: 1 & 0.139 & 0.222 & 0.222 & 0.222\\
PHQ9: 2 & 0.118 & 0.296 & 0.296 & 0.296\\
PHQ9: 3 & 0.450 & 0.548 & 0.548 & 0.548\\
Age at initiation squared & 0.000 & 0.001 & 0.001 & 0.001\\
Charlson score 1 \& anxiety disorder & -0.090 & -0.180 & -0.180 & -0.180\\
Charlson score 2 \& anxiety disorder & 0.298 & 0.146 & 0.146 & 0.146\\
Charlson score 3+ \& anxiety disorder & 0.033 & 0.260 & 0.260 & 0.260\\
Age at initiation \& female sex & 0.000 & -0.007 & -0.007 & -0.007\\
Female sex \& prior self-harm & 0.155 & -0.014 & -0.014 & -0.014\\
Age at initiation \& prior self-harm & -0.003 & -0.020 & -0.020 & -0.020\\
Charlson score 1 \& age at initiation & 0.002 & 0.002 & 0.002 & 0.002\\
Charlson score 2 \& age at initiation & -0.001 & -0.007 & -0.007 & -0.007\\
Charlson score 3+ \& age at initiation & -0.013 & -0.019 & -0.019 & -0.019\\
PHQ item 9 score 1 \& female sex & 0.085 & -0.042 & -0.042 & -0.042\\
PHQ item 9 score 2 \& female sex & 0.051 & -0.064 & -0.064 & -0.064\\
PHQ item 9 score 3 \& female sex & 0.026 & 0.059 & 0.059 & 0.059\\
PHQ item 9 score 1 \& prior self-harm & 0.497 & -0.218 & -0.218 & -0.218\\
PHQ item 9 score 2 \& prior self-harm & 0.889 & -0.494 & -0.494 & -0.494\\
PHQ item 9 score 3 \& prior self-harm & 0.330 & -0.534 & -0.534 & -0.534\\
\bottomrule
\end{tabular}
\end{table}

\clearpage

We generated the new source datasets of size $n = 10,000$ as follows: for each DGM, 
\begin{enumerate}
    \item sample covariates $W$ with replacement from the original KPWA source dataset;
    \item generate treatment assignments (psychotherapy = 1) conditioned on the sampled $W$ using a logistic regression model with coefficients given in Table~\ref{tab:original_coefficients} column 2;
    \item generate outcomes conditioned on $(A,W)$ using a logistic regression model with coefficients given in Table~\ref{tab:original_coefficients} column 4 (for the 15\% outcome) and column 5 (for the 5\% outcome).
\end{enumerate}

\section{Supplemental Tables and Figures for the electronic health records-based plasmode simulations}

\begin{table}[!htbp]
\centering
\caption{Results of the KPWA-based plasmode simulations for the 15\% outcome (ATE and RR).\label{tab:kpwa_plasmode_r0.15}}
\centering
\resizebox{\ifdim\width>\linewidth\linewidth\else\width\fi}{!}{
\fontsize{11}{13}\selectfont
\begin{threeparttable}
\begin{tabular}[t]{llllllrllllr}
\toprule
\multicolumn{2}{c}{ } & \multicolumn{5}{c}{Sample Treatment} & \multicolumn{5}{c}{Generate Treatment} \\
\cmidrule(l{3pt}r{3pt}){3-7} \cmidrule(l{3pt}r{3pt}){8-12}
Estimand & Estimator & \% Bias & SE & RMSE & bias:SE & CP & \% Bias & SE & RMSE & bias:SE & CP\\
\midrule
 & Unadj & 25.054 & 0.007 & 0.024 & 3.393 & 7.6 & 25.230 & 0.007 & 0.024 & 3.396 & 7.6\\
\cmidrule{2-12}
 & Match & 1.275 & 0.009 & 0.009 & 0.131 & 94.8 & 1.344 & 0.008 & 0.008 & 0.150 & 94.8\\
\cmidrule{2-12}
 & IPTW & -0.344 & 0.007 & 0.007 & 0.043 & 95.2 & -0.005 & 0.007 & 0.007 & $<$ 0.001 & 95.0\\
\cmidrule{2-12}
 & TMLE & -0.071 & 0.007 & 0.007 & 0.009 & 95.0 & 0.025 & 0.007 & 0.007 & 0.003 & 95.0\\
\cmidrule{2-12}
 & glmCM & -0.074 & 0.007 & 0.007 & 0.010 & 95.1 & -0.028 & 0.007 & 0.007 & 0.004 & 95.1\\
\cmidrule{2-12}
\multirow{-6}{*}{\raggedright\arraybackslash ATE} & glmPS & -1.348 & 0.007 & 0.007 & 0.180 & 94.6 & -1.316 & 0.007 & 0.007 & 0.175 & 94.7\\
\cmidrule{1-12}
 & Unadj & -16.171 & 0.024 & 0.086 & 3.512 & 6.3 & -16.299 & 0.024 & 0.087 & 3.518 & 6.3\\
\cmidrule{2-12}
 & Match & -1.421 & 0.037 & 0.038 & 0.196 & 94.8 & -1.353 & 0.034 & 0.035 & 0.202 & 94.7\\
\cmidrule{2-12}
 & IPTW & 0.491 & 0.031 & 0.031 & 0.082 & 95.1 & 0.044 & 0.030 & 0.030 & 0.008 & 95.0\\
\cmidrule{2-12}
 & TMLE & 0.127 & 0.030 & 0.030 & 0.022 & 95.0 & 0.069 & 0.030 & 0.030 & 0.012 & 95.1\\
\cmidrule{2-12}
 & glmCM & 0.137 & 0.028 & 0.028 & 0.025 & 95.0 & 0.105 & 0.028 & 0.028 & 0.019 & 95.0\\
\cmidrule{2-12}
\multirow{-6}{*}{\raggedright\arraybackslash RR} & glmPS & 0.623 & 0.028 & 0.028 & 0.113 & 94.9 & 0.579 & 0.028 & 0.029 & 0.104 & 94.8\\
\bottomrule
\end{tabular}
\begin{tablenotes}
\item Abbreviations: KPWA: Kaiser Permanente Washington; ATE: average treatment effect; RR: relative risk; SE: standard error; RMSE: root mean squared error; Bias:SE: ratio of bias to standard error; CP: coverage probability; Unadj: unadjusted; Match: propensity score matching; IPTW: inverse probability of treatment weighting; TMLE: targeted maximum likelihood estimation; glmCM: generalized linear model, correctly specified; glmPS: generalized linear model, adjusted for propensity score. Results for n = 10,000 ($1/\sqrt{n}=$0.01) with 100,000 Monte-Carlo iterations. The true value of the ATE is -0.092; the true value of the RR is 0.512.
\end{tablenotes}
\end{threeparttable}}
\end{table}

\clearpage



\begin{table}[!htbp]
\centering
\caption{Results of the KPWA-based plasmode simulations for the 15\% outcome (EY1, EY0, and logcOR).\label{tab:kpwa_plasmode_r0.15_supp}}
\centering
\resizebox{\ifdim\width>\linewidth\linewidth\else\width\fi}{!}{
\fontsize{11}{13}\selectfont
\begin{threeparttable}
\begin{tabular}[t]{llllllrllllr}
\toprule
\multicolumn{2}{c}{ } & \multicolumn{5}{c}{Sample Treatment} & \multicolumn{5}{c}{Generate Treatment} \\
\cmidrule(l{3pt}r{3pt}){3-7} \cmidrule(l{3pt}r{3pt}){8-12}
Estimand & Estimator & \% Bias & SE & RMSE & bias:SE & CP & \% Bias & SE & RMSE & bias:SE & CP\\
\midrule
 & Unadj & 6.840 & 0.005 & 0.014 & 2.394 & 33.4 & 6.867 & 0.005 & 0.014 & 2.394 & 33.4\\
\cmidrule{2-12}
 & Match & -0.333 & 0.007 & 0.007 & 0.096 & 95.0 & -0.178 & 0.006 & 0.006 & 0.055 & 95.0\\
\cmidrule{2-12}
 & IPTW & 0.088 & 0.006 & 0.006 & 0.030 & 95.1 & -0.042 & 0.006 & 0.006 & 0.014 & 95.0\\
\cmidrule{2-12}
 & TMLE & -0.020 & 0.006 & 0.006 & 0.007 & 95.0 & 0.013 & 0.006 & 0.006 & 0.005 & 95.0\\
\cmidrule{2-12}
 & glmCM & -0.013 & 0.005 & 0.005 & 0.005 & 95.0 & -0.002 & 0.005 & 0.005 & $<$ 0.001 & 95.0\\
\cmidrule{2-12}
\multirow{-6}{*}{\raggedright\arraybackslash EY0} & glmPS & -0.778 & 0.005 & 0.005 & 0.285 & 94.2 & -0.794 & 0.005 & 0.005 & 0.290 & 94.1\\
\cmidrule{1-12}
 & Unadj & -10.500 & 0.004 & 0.011 & 2.442 & 31.4 & -10.616 & 0.004 & 0.011 & 2.456 & 30.8\\
\cmidrule{2-12}
 & Match & -1.864 & 0.006 & 0.007 & 0.289 & 94.2 & -1.627 & 0.006 & 0.006 & 0.274 & 94.1\\
\cmidrule{2-12}
 & IPTW & 0.499 & 0.005 & 0.005 & 0.093 & 95.3 & -0.077 & 0.005 & 0.005 & 0.015 & 95.1\\
\cmidrule{2-12}
 & TMLE & 0.029 & 0.005 & 0.005 & 0.006 & 95.1 & 0.003 & 0.005 & 0.005 & $<$ 0.001 & 95.0\\
\cmidrule{2-12}
 & glmCM & 0.046 & 0.005 & 0.005 & 0.010 & 95.0 & 0.024 & 0.005 & 0.005 & 0.005 & 95.1\\
\cmidrule{2-12}
\multirow{-6}{*}{\raggedright\arraybackslash EY1} & glmPS & -0.236 & 0.005 & 0.005 & 0.050 & 95.0 & -0.297 & 0.005 & 0.005 & 0.062 & 95.0\\
\cmidrule{1-12}
logcOR & glmCM & 0.505 & 0.069 & 0.069 & 0.063 & 95.0 & 0.555 & 0.070 & 0.070 & 0.068 & 94.9\\
\bottomrule
\end{tabular}
\begin{tablenotes}
\item Abbreviations: KPWA: Kaiser Permanente Washington; EY0: expected value of outcome under no treatment; EY1: expected value of outcome under treatment; logcOR: conditional log odds ratio; SE: standard error; RMSE: root mean squared error; Bias:SE: ratio of bias to standard error; CP: coverage probability; Unadj: unadjusted; Match: propensity score matching; IPTW: inverse probability of treatment weighting; TMLE: targeted maximum likelihood estimation; glmCM: generalized linear model, correctly specified; glmPS: generalized linear model, adjusted for propensity score. Results for n = 10,000 ($1/\sqrt{n}=$0.01) with 100,000 Monte-Carlo iterations. The true value of EY0 is 0.19; the true value of EY1 is 0.097; the true value of logcOR is -0.861.
\end{tablenotes}
\end{threeparttable}}
\end{table}

\clearpage

\begin{table}[!htbp]
\centering
\caption{Results of the KPWA-based plasmode simulations for the 5\% outcome (EY1, EY0, and logcOR).\label{tab:kpwa_plasmode_r0.05_supp}}
\centering
\resizebox{\ifdim\width>\linewidth\linewidth\else\width\fi}{!}{
\fontsize{11}{13}\selectfont
\begin{threeparttable}
\begin{tabular}[t]{llllllrllllr}
\toprule
\multicolumn{2}{c}{ } & \multicolumn{5}{c}{Sample Treatment} & \multicolumn{5}{c}{Generate Treatment} \\
\cmidrule(l{3pt}r{3pt}){3-7} \cmidrule(l{3pt}r{3pt}){8-12}
Estimand & Estimator & \% Bias & SE & RMSE & bias:SE & CP & \% Bias & SE & RMSE & bias:SE & CP\\
\midrule
 & Unadj & 9.457 & 0.004 & 0.009 & 2.022 & 47.9 & 9.659 & 0.004 & 0.009 & 2.060 & 46.3\\
\cmidrule{2-12}
 & Match & 0.137 & 0.005 & 0.005 & 0.024 & 95.0 & -0.365 & 0.004 & 0.004 & 0.071 & 95.0\\
\cmidrule{2-12}
 & IPTW & 1.193 & 0.005 & 0.005 & 0.216 & 95.3 & -0.210 & 0.004 & 0.004 & 0.045 & 95.1\\
\cmidrule{2-12}
 & TMLE & 0.517 & 0.005 & 0.005 & 0.096 & 95.2 & 0.002 & 0.004 & 0.004 & $<$ 0.001 & 95.1\\
\cmidrule{2-12}
 & glmCM & -0.166 & 0.004 & 0.004 & 0.036 & 95.3 & -0.176 & 0.004 & 0.004 & 0.039 & 95.2\\
\cmidrule{2-12}
\multirow{-6}{*}{\raggedright\arraybackslash EY0} & glmPS & -2.264 & 0.004 & 0.004 & 0.516 & 92.0 & -2.528 & 0.004 & 0.004 & 0.577 & 91.0\\
\cmidrule{1-12}
 & Unadj & -13.214 & $<$ 0.001 & 0.001 & 0.701 & 88.5 & -13.394 & $<$ 0.001 & 0.001 & 0.711 & 88.4\\
\cmidrule{2-12}
 & Match & -3.866 & 0.002 & 0.002 & 0.130 & 95.6 & -2.165 & 0.001 & 0.001 & 0.077 & 95.5\\
\cmidrule{2-12}
 & IPTW & 1.255 & 0.001 & 0.001 & 0.050 & 95.2 & -0.077 & 0.001 & 0.001 & 0.003 & 95.2\\
\cmidrule{2-12}
 & TMLE & -0.303 & 0.001 & 0.001 & 0.012 & 95.2 & -0.144 & 0.001 & 0.001 & 0.006 & 95.2\\
\cmidrule{2-12}
 & glmCM & -0.031 & 0.001 & 0.001 & 0.001 & 95.1 & -0.083 & 0.001 & 0.001 & 0.004 & 95.1\\
\cmidrule{2-12}
\multirow{-6}{*}{\raggedright\arraybackslash EY1} & glmPS & 2.088 & 0.001 & 0.001 & 0.094 & 95.0 & 2.675 & 0.001 & 0.001 & 0.120 & 94.8\\
\cmidrule{1-12}
logcOR & glmCM & 1.824 & 0.247 & 0.253 & 0.222 & 94.3 & 1.790 & 0.247 & 0.253 & 0.218 & 94.3\\
\bottomrule
\end{tabular}
\begin{tablenotes}
\item Abbreviations: KPWA: Kaiser Permanente Washington; EY0: expected value of outcome under no treatment; EY1: expected value of outcome under treatment; logcOR: conditional log odds ratio; SE: standard error; RMSE: root mean squared error; Bias:SE: ratio of bias to standard error; CP: coverage probability; Unadj: unadjusted; Match: propensity score matching; IPTW: inverse probability of treatment weighting; TMLE: targeted maximum likelihood estimation; glmCM: generalized linear model, correctly specified; glmPS: generalized linear model, adjusted for propensity score. Results for n = 10,000 ($1/\sqrt{n}=$0.01) with 100,000 Monte-Carlo iterations. The true value of EY0 is 0.084; the true value of EY1 is 0.005; the true value of logcOR is -3.013.
\end{tablenotes}
\end{threeparttable}}
\end{table}

\clearpage

\fi

\bibliographystyle{apalike} 
\bibliography{PlasmodeSimulation}

\end{document}